\documentclass{article}
\usepackage{epsfig,amsmath,amssymb,euscript}

\makeatletter

\newcommand{\B}[1]{\mathop{\mbox{\boldmath$ #1 $}}\nolimits}
\newtheorem{theorem}{Theorem}
\newtheorem{remark}{Remark}
\newtheorem{corollary}{Corollary}
\newtheorem{lemma}{Lemma}
\newtheorem{proposition}{Proposition}

\begin{document}

$\ $

\vspace{20pt}
\centerline{\large\bf EFFICIENT NONPARAMETRIC ESTIMATION AND }
\vspace{2pt}
\centerline{\large\bf INFERENCE FOR THE VOLATILITY FUNCTION}
\vspace{20pt}
\centerline{Francesco Giordano\footnote{Department of Economics and Statistics, University of Salerno - Via Giovanni Paolo II n.137 - 84084 Fisciano (SA) - Italy, giordano@unisa.it} and Maria Lucia Parrella\footnote{Department of Economics and Statistics, University of Salerno - Via Giovanni Paolo II n.137 - 84084 Fisciano (SA) - Italy, mparrella@unisa.it, tel:+39 089962211, fax:+39 089962049}}
\vspace{20pt}

%%%%%%%%%%%%%%%%%%%%%%%%%%%%%%%%%%%%%%%%%%%%%%%%%%%%%%%%%%%%%%%%%%%%%%%%%%%%%%%%%%%%%%%%%%%%%%%%%%%%%%%%%%%%%%%%%%%%%%%%%%%%

\begin{quotation}
\noindent {\textbf{Abstract}:}
During the last decades there has been increasing interest in modeling the volatility of financial data. Several parametric models have been proposed to this aim, starting from ARCH, GARCH and their variants, but often it is hard to evaluate which one is the most suitable for the analyzed financial data. In this paper we focus on nonparametric analysis of the volatility function for mixing processes. Our approach encompasses many parametric frameworks and supplies several tools which can be used to give evidence against or in favor of a specific parametric model: nonparametric function estimation, confidence bands and test for symmetry. Another contribution of this paper is to give an alternative representation of the $GARCH(1,1)$ model in terms of a \emph{Nonparametric-ARCH(1)} model, which avoids the use of the lagged volatility, so that a more precise and more informative  \emph{News Impact Function} can be estimated by our procedure.
We prove the consistency of the proposed method and investigate its empirical performance on synthetic and real datasets. Surprisingly,  for finite sample size, the simulation results show a better performance of our nonparametric estimator compared with the MLE estimator of a $GARCH(1,1)$ model, even in the case of correct specification of the model.\par

\vspace{9pt}
\noindent {\it \textbf{Key words and phrases:}}
Nonparametric volatility estimation, confidence intervals for volatility, testing for symmetry.
\par
\end{quotation}\par

\section{Introduction} 

The importance of a correct specification for volatility models has been confirmed since the work of \cite{EngNg93}. Several attempts have been made from then to deal with the volatility processes nonparametrically, in order to avoid the mispecification problems and to produce robust estimation results. There have been different approaches that focus (alternatively) on the error density, on the functional form of the volatility function, or the kind of nonparametric estimator. See, among others, \cite{FanYao98, HarTsy97, FraDia06, XuPhi11, WanAlt12, HarAlt15}.

First of all, \cite{HarTsy97} proposed to estimate the $ARCH(p)$ class of models using the local linear estimator, where $p$ is the number of lags in the model. But their model suffers from the well-known \emph{curse of dimensionality} problem which affects the nonparametric estimators. In fact, the best rate of convergence of any nonparametric estimator of a function is $n^{-2/(4+p)}$, where $n$ is the time series length and $p$ the number of covariates (=lags) in the model \cite{Gyorfy02}. This rate is extremely slow when $p$ is large. Therefore, \cite{AudBul01} and \cite{Bulmcn02} proposed a nonparametric procedure based on a bivariate smoother in order to nest the $GARCH(1,1)$ class of models of \cite{Bol86}, which are known to be equivalent to the $ARCH(\infty)$ although they only need two covariates. Therefore, the convergence rate improves from $n^{-2/(4+p)}$ to $n^{-1/3}$. However, the difficulties with the proposal of \cite{Bulmcn02} are given by i) the initialization of the latent process used as a covariate in the $GARCH$ smoother and ii) the choice of the bivariate bandwidth (tuning parameter). A different (semi-parametric) approach has been recently proposed by \cite{WanAlt12}, where a GARCH(1,1) model is approximated by a truncated additive model with nonparametric components, estimated by smoothing splines and linked together by a common parametric coefficient. However, also in the last paper, the problem of bandwidth selection still remains a crucial and unsolved issue.

Although the many proposals, nonparametric methods for volatility analysis have not gained much interest from practitioners and, therefore, research on such approaches has not increased in recent years, contrary to what has happened with parametric approaches. The main reasons have been: i) the difficulty to set the tuning parameters of the nonparametric procedures and ii) the slow convergence rate of the nonparametric estimators. These two drawbacks have not been sufficiently compensated by the gain in robustness of the nonparametric analysis.

Trying to deal with the two drawbacks, in this paper we show that nonparametric methods can give important and essential contributions to financial data analysis, mainly from an inferential point of view. In fact, risk evaluation and volatility forecasts are some of the goals that require the selection of a suitable parametric model for the data generating process in order to get consistent results. So, in order to validate analysis using empirical evidence, parametric estimators should be replaced by nonparametric ones, or at least compared with them basing on nonparametric confidence intervals and/or tests for symmetry.
Therefore, in this paper, we focus on a general nonparametric framework for the analysis of the volatility function. Our aim is to provide a set of tools that can be used for robust volatility analysis: \emph{a}) a consistent nonparametric estimator of the volatility function based on local smoothing with data-driven optimal bandwidth, \emph{b}) the nonparametric confidence intervals for volatility model selection and \emph{c}) a nonparametric test for symmetry of the volatility function. All this is made by means of a Nonparametric Autoregressive Conditional Heteroskedastic model of order one, here denoted as $NARCH(1)$. 

Now we summarize how we avoid the two drawbacks above. To face the problem of setting the tuning parameters, we extend the smoothing procedure of \cite{GioPar14} to the case of heteroskedastic and autoregressive models. This method is based on a hybrid, data-driven bandwidth estimator, a cross between local and global smoothing, which encompasses the adaptability of local smoothing and the efficiency of global smoothing. As a result, it allows to reach the best rate of convergence for the final nonparametric volatility estimator. Note that we cannot directly apply the theoretical results in \cite{GioPar14} because here we have dependent data.
To deal with the second problem (improving the convergence rate of the nonparametric volatility estimator), we reduce the number of regressors in the model. In fact, remembering that the best rate of convergence of any nonparametric estimator of a function is $n^{-2/(4+p)}$, we show that the nonlinear and nonadditive structure of the $NARCH(1)$ model can be exploited in order to capture the dependence of the process by means of only one regressor (\emph{i.e.}, $p=1$) for many parametric models. As an application of this idea, we show in section \ref{MOD} how the $GARCH(1,1)$ model can be equivalently represented by a particular $NARCH(1)$ model, with two immediate advantages: i) we avoid the (latent) lagged volatility as a covariate of the model, so that we can build a more precise \emph{News Impact Curve}, which can be used to give effective evidence of \emph{leverage effects} in the data; ii) we improve the convergence rate of the nonparametric volatility estimator from $n^{-1/3}$ (reached by \cite{Bulmcn02}) to $n^{-2/5}$ (reached by our nonparametric estimator). All this may have a strong positive impact on the efficiency of volatility estimates, as shown by simulation results.

The rest of the paper is organized as follows. Section \ref{MOD} presents the nonparametric model and gives the main idea concerning the new representation of the $GARCH(1,1)$ model and the new interpretation of the \emph{News Impact Curve}. Section \ref{idea} describes the estimation procedure and gives the theoretical results on consistency both for the asymptotic optimal bandwidth and volatility function estimators. The derivation of the confidence intervals and the test for symmetry are shown in sections \ref{Conf} and \ref{Test}, respectively. The empirical performance of the method is investigated in section \ref{sim}, with simulated data, and section \ref{data}, with a real dataset. Some concluding remarks are given in section \ref{final}. All the assumptions and the proofs are concentrated in the Appendix.

\section{An adaptive nonparametric setup for volatility}\label{MOD}

Consider a stationary process $\left\{X_t\right\}$ and define a \emph{Nonparametric Autoregressive Conditional Heteroscedastic}  model of order 1, the $NARCH(1)$, as follows
\begin{equation}\label{eqn01}
 X_{t}=\sigma(X_{t-1})\varepsilon_{t}, \qquad\qquad t\in\mathbb{N},
\end{equation}
where the errors $\varepsilon_t$ are independent and identically distributed real random variables, satisfying $E(\varepsilon_t)=0$ and
$\mathop{Var}(\varepsilon_t)=1$ for each $t$. For simplicity, we assume that the conditional mean function $m(x)=E\left\{X_{t}|X_{t-1}=x\right\}$ is equal to zero. This setup is typically considered when analyzing financial data, where no conditional structure in the mean is generally observed from data (otherwise, it is sufficient to work with the residual process $R_t=X_t-m(X_{t-1})$ as in \cite{FanYao98} and \cite{HarTsy97}). Here and in the sequel, $x$ represents a generic point of the support of $X_t$. Given model (\ref{eqn01}), we look at the conditional variance function, also known as \emph{volatility function},
\begin{equation}\label{mphi.x}
\sigma^2(x)=\mathop{Var}\left\{X_{t}|X_{t-1}=x\right\}.
\end{equation}
By (\ref{mphi.x}), we have a general class of volatility functions and the error term $\varepsilon_t$ is also general enough (see assumptions (a) in the Appendix), so that model (\ref{eqn01}) encompasses many parametric volatility models proposed in the literature. 
%The upper bound for the convergence rate of any nonparametric estimator is $\sigma^2(x)-\hat\sigma^2(x)=O_p(n^{-2/(4+1)})$, as shown by ???
In particular, it is immediate to see that the classic $ARCH(1)$ model is a particular case of model (\ref{eqn01}), given by the linear equation $\sigma^2(X_{t-1})\equiv \alpha_0+\alpha_1X_{t-1}^2$, with $\alpha_i>0$, $i=0,1$, and $\alpha_1<1$. Other examples are the generalizations of $ARCH$ models, such as the threshold based $TARCH(1)$.
 
The $ARCH(1)$ model and its variants are often accused to perform poor with real data. In practice, one needs many lagged variables in the model to match the dependence found in financial data, which implies the need of $ARCH(p)$ models, where $\sigma^2(X_{t-1},\ldots,X_{t-p})=\alpha_0+\alpha_1X^2_{t-1}\ldots+\alpha_pX_{t-p}^2$. The estimation of such models can be inefficient when $p$ is large. This has motivated the orientation towards the $GARCH(1,1)$ model, which is one of the most used in financial econometrics. It is given by
\begin{eqnarray}
X_{t} &=& \sigma_{t}\varepsilon_{t} \label{garch}\\
\sigma_t^2 &=& \alpha_0+\alpha_1X^2_{t-1}+\beta \sigma_{t-1}^2, \nonumber
\end{eqnarray}
with $\alpha_i>0$, $i=0,1$, $\beta\ge 0$ and $\alpha_1+\beta<1$. The advantage with this model is that it is formally equivalent to the $ARCH(\infty)$, although the dependence structure is captured by only two regressors ($X_{t-1}$ and $\sigma_{t-1}$) instead of infinite regressors. Several studies have established the good performance of the $GARCH(1,1)$ model compared to $GARCH(p,q)$ and to many other volatility models (see, for example, \cite{HanLun01}). But a serious problem is given by the fact that the regressor $\sigma_{t-1}$ is a latent process (the lag of volatility itself) which must be estimated or substituted by some reliable proxy. As a consequence, the estimation of the $GARCH(1,1)$ model (and its variants) is not trivial and may lead to unstable results.

In this section, we show that the classic $GARCH(1,1)$ model can be equivalently represented as a \emph{nonparametric $ARCH(1)$ model}, that is the $NARCH(1)$ defined in (\ref{eqn01}). The advantage of this new representation is threefold: \emph{a}) the new model is able to capture the dependence structure of a $GARCH(1,1)$, and therefore of an $ARCH(\infty)$, by means of only one covariate; \emph{b}) such a covariate is the lag $X_{t-1}$, which is an observed process; \emph{c}) a different and more precise \emph{News Impact Curve} can be derived and estimated for the new model. This threefold advantage is obtained thanks to the nonparametric structure of the model, which allows to capture the effects of the infinite lags $X_{t-j}$ on the volatility by means of the adaptive and nonlinear structure of the volatility function itself. 
In other words, we allow the function $\sigma(\cdot)$ to be ``free''  and therefore ``capable'' to well suit the relation between $X_t$ and its past, comdensed in $X_{t-1}$.
This is stated in Theorem \ref{theorem1}.

\begin{theorem}\label{theorem1}
Assuming a symmetric density for the error $\varepsilon_t$, the $GARCH(1,1)$ model in (\ref{garch}), with parameters $\alpha_0>0, \alpha_1>0$, $\beta\ge 0$ and $\alpha_1+\beta<1$, is equivalent to a nonlinear volatility model as in (\ref{eqn01}), where the volatility function $\sigma^2(x)$ is given by

\begin{eqnarray}\label{sigmax}
\sigma^2(x) &=&\left\{\begin{array}{lll}
A_0&\mbox{if}&x=0,\\
A_0+(\alpha_1+\beta)\tilde g(x;\alpha_1,\beta)x^2&\mbox{if}&x\not =0, 
\end{array}
\right.
\end{eqnarray}

where
\[
\tilde g(x;\alpha_1,\beta)=g(x;\alpha_1,\beta)-\frac{B_0}{x^2(\alpha_1+\beta)}
\]
with
\[
g(x;\alpha_1,\beta)\equiv E\left(\left.\frac{1}{C_{\tilde{\varepsilon}_t}}\right|X_{t}=x\right), \quad B_0=\beta\alpha_0/(1-\alpha_1-\beta),  \quad A_0=\alpha_0+B_0, 
\]
\[C_{\tilde{\varepsilon}_t}=1+\beta/\alpha_1(1-1/\tilde{\varepsilon}_t^2)\quad \mbox{ and }\quad \tilde\varepsilon_t=\mathop{sign}(\varepsilon_t)\sqrt{\frac{\alpha_1\varepsilon^2_t+\beta}{\alpha_1+\beta}}.
\]
\end{theorem}
\begin{remark}
\label{rem1bis}
Theorem \ref{theorem1} can be generalized in two directions. First, we can relax the assumption of symmetry for the density function of $\varepsilon_t$. We only use it to simplify the proof of Theorem \ref{theorem1} in order to derive that $E\left(\tilde{\varepsilon}_t\right)=0$. Second, we can extend the result of Theorem \ref{theorem1} to nonparametric \textit{GARCH}(1,1) models.
\end{remark}

By Theorem \ref{theorem1} we have that the \textit{GARCH}(1,1) process can be written as $X_t=C_{\tilde{\varepsilon}_t}^{1/2}\tilde{X}_t$ where $\tilde X_t\sim ARCH(1;\alpha_0,\alpha_1+\beta)$, with the error terms $\tilde{\varepsilon}_t$. Note that this representation is exact. Of course, $X_t$ is not an \textit{ARCH}(1) process, since one can show that $E\left[(X_t-\tilde X_t)^2\right]>0$, $\forall t$. However, $E(X_t^2)=E(\tilde X_t^2)$ and this is used in Theorem \ref{theorem1} to show that the \textit{GARCH}(1,1) process can be equivalently represented by a particular $NARCH(1)$ structure, which only depends on $X_{t-1}$. In fact, for a given value of $X_{t-1}=x$, the volatility function is $\sigma^2(x)=A_0+(\alpha_1+\beta)\tilde g(x;\alpha_1,\beta)x^2\equiv A_0+\tilde\alpha_1(x)x^2$, that represents a ``rescaled'' $ARCH(1)$ model with support-dependent coefficient $\tilde\alpha_1$. 
To summarize, Theorem \ref{theorem1} shows that there exists a \textit{nonlinear} representation of the volatility function from a \textit{GARCH}(1,1) process which only depends on $X_{t-1}$ instead of its classic \textit{linear} representation with infinite variables, $ARCH(\infty)$.

It is important to stress that the value of the coefficient $\tilde\alpha_1(x)$ in model (\ref{sigmax}) changes with $x$, so that we have a function of coefficients instead of a single coefficient to estimate. However, this function of coefficients \textbf{cannot} be expressed in closed form, therefore model (\ref{sigmax}) \textbf{cannot} be analyzed and estimated with parametric methods.  In fact, it would be necessary to know the density of the errors $\varepsilon_t$ in order to derive the analytic form of $\tilde g(x;\cdot)$, but even in the simplest case such derivations would be difficult (and the parametric estimation methods, such as maximum likelihood or quasi-maximum likelihood, are impossible to apply). Therefore, Theorem \ref{theorem1} is useless in the parametric framework, but it has a natural application in the nonparametric framework.
In fact, note that we do not need to compute  (explicitly)  the component $\tilde g(x;\cdot)$ in order to make inference or generate predictions. We just need to guarantee that the estimation procedure is able to incorporate this component in the final estimations. The nonparametric procedure proposed in section \ref{idea}, based on the local polynomial estimator with optimal data-driven local bandwidth, perceives such a goal. 

\begin{figure}[t]
    \centering
 \resizebox{10 cm}{!}{\includegraphics{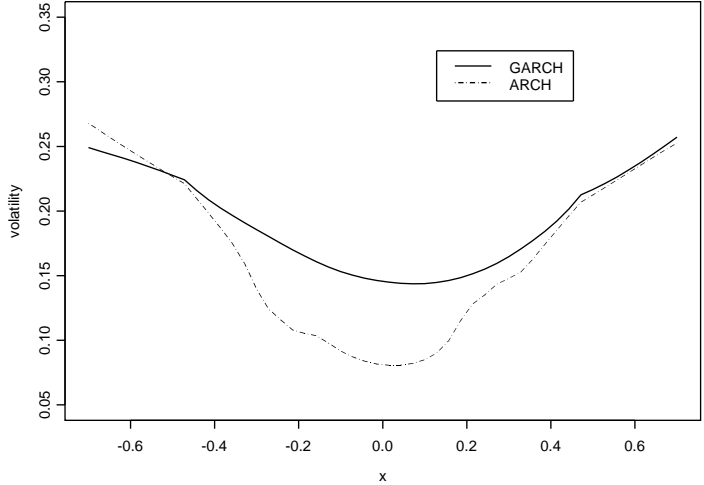}}
    \caption{\footnotesize The \emph{News Impact Curve} estimated nonparametrically on two different datasets. The first one is generated from a \textit{GARCH}(1,1;0.1,0.3,0.2) with standard normal error term. The second one is generated from an \textit{ARCH}(1;0.1,0.5) with the error term $\widetilde{\varepsilon}_t$ defined in Theorem \ref{theorem1}. The difference between the two curves reflects the component $\tilde g(x;\cdot)$ defined in Theorem \ref{theorem1}.}
    \label{figmG}
\end{figure}

\subsection{A new interpretation of the News Impact Curve}
Theorem \ref{theorem1} has important consequences for the interpretation of the \emph{News Impact Curve (NIC)}. The \emph{NIC} has been first defined by \cite{EngNg93} for $GARCH$ models and its variants, to measure how new information is incorporated into volatility estimates. It is defined as the implied relation between $X_{t-1}$ and $\sigma_t^2$, once considered constant the information at time $t-2$ and earlier, so that $\sigma_t^2=\alpha_0+\alpha_1X_{t-1}^2+\beta\alpha_0/(1-\alpha_1-\beta)$ (see \cite{EngNg93}, p. 1754). In practice, the \emph{NIC} is derived by imposing the lagged volatility value $\sigma^2_{t-1}$ to be equal to its unconditional mean $\beta\alpha_0/(1-\alpha_1-\beta)$. This choice (conditioning to the unconditional mean) is strictly necessary in order to draw the \emph{NIC} as a function of the $X_{t-1}$ alone, so that it can be plotted as the well-known $U$-shaped curve. The main utility of such a curve is to give evidence of \emph{leverage effects} in the data.

Now, by (\ref{sigmax}) in Theorem \ref{theorem1}, the volatility function of a $GARCH(1,1)$ model can be reformulated as a nonlinear function of the lagged return $X_{t-1}$ alone. So, it is not necessary to set a value for the lagged volatility in order to plot the function, although the effect of the lagged volatility is incorporated in the NIC by means of $\tilde g(x;\cdot)$.
%Therefore, a new representation of the \emph{NIC} function can be built to better capture the relation between the return $X_{t-1}$ and the volatility $\sigma_t^2$. 
In other words, instead of using the constant $\beta\alpha_0/(1-\alpha_1-\beta)$, we take advantage of the function $\tilde g(x;\cdot)$ to improve local adaptivity of the NIC. 

Figure \ref{figmG} gives an illustrative example of the \emph{NIC} for two different models and also shows empirically the result of Theorem \ref{theorem1}. We report the volatility function $E(X_t^2|X_{t-1}=x)$ estimated nonparametrically on two different datasets, using the procedure suggested in section \ref{idea}. The first dataset is generated from a $GARCH(1,1)$ model, with $\alpha_0=0.1,\alpha_1=0.3,\beta=0.2$ and standard normal error $\varepsilon_t$. The second dataset originates from an $ARCH(1)$ model with $\alpha_0=0.1,\tilde\alpha_1=\alpha_1+\beta=0.5$ and error term $\widetilde{\varepsilon}_t$ defined in Theorem \ref{theorem1}. 
From a theoretical point of view, as shown in Theorem \ref{theorem1}, the two volatility models are not equivalent because $\tilde\alpha_1$ is constant in the $ARCH$ model. As a consequence, the two curves in Figure \ref{figmG} do not coincide and the difference between them reflects the component $\tilde g(x;\cdot)$ defined in Theorem \ref{theorem1} (plus a constant term). Note that the two functions represent the \emph{NIC} for the two models.
We can observe that they tend to have the same behaviour for large values of $|x|$ whereas the \textit{NIC} of the $GARCH(1,1)$ shows an inflation of the volatility function with respect to the $ARCH(1)$ case for small values of $|x|$. In fact, by Theorem \ref{theorem1}, the $GARCH(1,1)$ curve has a minimum at zero which is $A_0$. Instead, $ARCH(1)$ curve exhibits the minimum at the same point but with value $\alpha_0$. 
  
\section{Nonparametric estimation of volatility}\label{idea}
For the estimation of the volatility function we generalize the global adaptive smoothing procedure (GAS) proposed by \cite{GioPar14}. In the appendix we give the theoretical results which extend the consistency of GAS to the current setup of $\alpha$-mixing processes. 
%his is not straightforward since we need to introduce, both in the procedure and in the proofs, some technical elements linked to the heteroscedastic autoregressive process. 

Given a realization of the process $\{X_t;t=1,\ldots,n\}$, the volatility function $\sigma^2(x)$ is estimated using a local linear estimator (LLE) with adaptive bandwidth function.
Let $K:[-1,1]\rightarrow\mathbb{R}$ be a density function, henceforth called \emph{kernel}, and write $\sigma^2_{(2)}(\cdot)$ for the derivative of order 2 of the volatility function. Assuming that $\sigma_{(2)}^2(\cdot)$ exists at the point $x\in\mathbb{R}$, the LLE of $\sigma^2(x)$ can be written as a weighted linear estimator
\begin{equation}\label{phi.hat}
\widehat{ \sigma}^2(x;h) = \sum_{t=2}^n X^2_{t} W_{K,h}\!\left(X_{t-1}-x\right),
\end{equation}
where $h$ is the bandwidth and $W_{K,h}(\cdot)$ gives the \emph{effective kernel weights}. These weights are derived by locally approximating the function with a line. Local linear estimators are well established and they are implemented in all statistical softwares. See, for example, the \texttt{KernSmooth} package for R; see also \cite{FanGij96} for further details on LLE.

As with all nonparametric methods, the crucial step with LLE is setting the bandwidth $h$, that behaves as a tuning parameter and affects the consistency of the nonparametric estimator. It may happen, with nonparametric and semi-parametric procedures, that tuning parameters are set by rule of thumb, given the difficulty of setting them automatically (see, for example, section 2.4 in \cite{WanAlt12}). In this paper we avoid this drawback and propose a self-contained data-driven method. To do this, we extend the approach of \cite{GioPar14} in order to deal with dependent data. In general, there are two categories of bandwidths: global (\emph{i.e.}, constant, not dependent on $x$) and local (\emph{i.e.} variable with $x$). The smoothing procedure proposed in \cite{GioPar14} is based on an hybrid, data-driven bandwidth estimator which exploits the advantages of both local smoothing (adaptability) and global smoothing (efficiency). This procedure has a better performance than other procedures (Cross-Validation, plug-in global smoothing) in terms of mean squared error, and reaches the optimal convergence rate of the final smoothing estimator $\hat\sigma^2(x;h)$, as shown in \cite{GioPar14}. Further simulation results, not reported here, Moreover, another advantage of the GAS procedure is that it exploits bandwidth estimation in order to derive all the pivotal quantities necessary to make inference for the volatility function. As a result, what is generally seen as a drawback of kernel regression (the necessity of estimating the bandwidth) becomes here the main tool to make inference on the estimated function. 
 
The method is as follows. Define a compact subset $I_x$, centered at the point $x$, such that $I_x=[x-a/2, x+a/2]$,
with $a>0$. The \emph{global adaptive bandwidth} is
\begin{equation}\label{h.i}
h_{I_x}=\left\{\frac{\mathbb{V}_{\omega_{I_x}}}{4n\mathbb{B}_{\omega_{I_x}}}\right\}^{1/5},
\end{equation}
where
\begin{equation}\label{fun1}
\mathbb{B}_{\omega_{I_x}}= C_1^2\int_{I_x}[\sigma^2_{(2)}(u)]^2f_X(u)d\omega_{I_x}(u),\quad \mathbb{V}_{\omega_{I_x}} = C_2\int_{I_x}V(u)d\omega_{I_x}(u),
\end{equation}
$f_X(\cdot)$ and ${\mu_X(\cdot)}$ are the density and the measure of the process, respectively, $d\omega_{I_x}(u)={du}/{\mu_X(I_x)}$, $V(x)=\mathop{Var}(X^2_{t}|X_{t-1}=x)$, while $C_1$ and $C_2$ are known constants depending on the kernel function. See \cite{GioPar14} for further details and an explanation on how to set the parameter $a$.

In the following, we propose the estimators of $\mathbb{B}_{\omega_{I_x}}$ and $\mathbb{V}_{\omega_{I_x}}$ in (\ref{fun1}), which can be plugged into the (\ref{h.i}) to obtain the bandwidth estimator $\widehat{ h}_{I_x}$. Note that such functionals are connected with the conditional bias and the conditional variance of the estimator (\ref{phi.hat}), respectively. Therefore, they will also be used in sections \ref{Conf} and \ref{Test} to derive the confidence intervals and to test the symmetry of the volatility function.

For $r\in \mathbb{N}$, let $m_r(x)$ be the conditional moment function $E(X_{t}^r|X_{t-1}=x)$. Then $\sigma^2(x)\equiv m_2(x)$ and $V(x) \equiv m_4(x)-m_2^2(x)$.
Generally, nonparametric estimation of $V(x)$ implies two separate estimations of $m_4(x)$ and $m_2(x)$, as in \cite{HarTsy97,FanYao98,FraDia06}. This approach is rather inefficient. To gain efficiency, we propose an alternative approach based on only one estimation. It uses the following reparameterization of model (\ref{eqn01})
\begin{equation}\label{riparam}
V(x) = m_4(x)-m_2^2(x) = m_2^2(x)\left(m_{4\varepsilon}-1\right),
\end{equation}
where $m_{4\varepsilon}=E(\varepsilon_t^4)$. Then we consider an estimator of $m_2(x)$, that is the Neural Networks one. Denote it by $q\!\left(x;\B{\eta}\right)$ and it is estimated by
\begin{equation}\label{approximator}
\B{\widehat{ \eta}}=\arg\min_{\B{\eta}}\sum_{t=2}^n
\left[X^2_{t}-q\!\left(X_{t-1};\B{\eta}\right)\right]^2.
\end{equation}
Now using (\ref{riparam}) and (\ref{approximator}), we propose the estimator $\widehat{ V}(x) = \widehat{ m}_2^2(x)\left[\widehat{m}_{4\varepsilon}-1\right]$, where
\begin{equation}\label{est.new1}
\widehat{ m}_2(x) \equiv q(x, \B{\widehat{ \eta}}), \quad \widehat{ m}_{4\varepsilon} =
\frac{\sum_{t=2}^{n} X_{t-1}^4}{\sum_{t=2}^{n} [q(X_{t-1}, \B{\widehat{ \eta}})]^2}.
\end{equation}
Next, we use again $q(x, \B{\widehat{ \eta}})$ to estimate the derivative $\sigma^2_{(2)}(x)$ by
\begin{equation}\label{est.new2}
\widehat{ \sigma}^2_{(2)}(x)\equiv q_{(2)}(x;\B{\widehat{\eta}}).
\end{equation}
Finally, the estimators for the functionals in (\ref{fun1}) are
\begin{eqnarray}\label{Rhat}
\widehat{\mathbb{B}}_{\omega_{I_x}} &=& \frac{C_1^2\sum_{t=2}^{n}\left[\widehat{\sigma}^2_{(2)}(X_{t-1})\right]^2\mathbb{I}(X_{t-1}\in
I_x)}{\sum_{t=2}^{n}\mathbb{I}(X_{t-1}\in I_x)} \\
\widehat{\mathbb{V}}_{\omega_{I_x}} &=& \frac{C_2\sum_{i=1}^{n^*}\widehat{V}(z_i)/n^*}{\sum_{t=2}^{n}\mathbb{I}(X_{t-1}\in
I_x)/n}, \nonumber
\end{eqnarray}
where $\mathbb{I}(\cdot)$ is the indicator function and the points $\{z_1, z_2, \ldots, z_{n^*}\}$ are values that are
equally spaced from the interval $I_x$, with $n^*=O(n)$.

Next, we consider the optimal bandwidth and its plug-in estimator
for the unknown function $\sigma^2(\cdot)$ in model (\ref{eqn01})
using the local linear estimator. So, we have the true bandwidth and its estimator, as 
\[
h_{I_x}=\left\{\frac{\mathbb{V}_{\omega_{I_x}}}{4n\mathbb{B}_{\omega_{I_x}}}\right\}^{1/5}\mbox{and}\quad
\widehat{h}_{I_x}=\left\{\frac{\widehat{\mathbb{V}}_{\omega_{I_x}}}{4n\widehat{\mathbb{B}}_{\omega_{I_x}}}\right\}^{1/5}.
\]
Let $I_x^n=[x-a_n/2,x+a_n/2]$ and $I_x=[x-a/2,x+a/2]$, where
$\{a_n\}$ is a bounded and positive sequence and $a>0$.  Instead, when $a_n\rightarrow 0$, it follows that $h_{I_x^n}\rightarrow h^{opt}(x)$ when $n\rightarrow\infty$, where $h^{opt}(x)$ is the local bandwidth given by 
\[
h^{opt}(x)=\left\{\frac{\mathcal{V}(x)}{4n\mathcal{B}^2(x)}\right\}^{1/5},
\]
where 
\begin{equation}
\label{fun2}
\mathcal{B}(x)=C_1\sigma_{(2)}^2(x)\quad \mbox{ and } \quad \mathcal{V}(x)=C_2\frac{V(x)}{f_X(x)}.
\end{equation}

We can state the following theorem.

\begin{theorem} \label{theorem2}
If the assumptions (a1) -- (a5) and (b1) -- (b3) hold and assume that
$[\sigma_{(2)}^2(x)]\not= 0$ in $I_x$, then $\widehat{h}_{I_x^n}$ is
consistent in the sense that:
\begin{itemize}

\item \textit{if $Q_1(n)\rightarrow 0$ as $n\rightarrow \infty$, then}
\begin{eqnarray*}
\frac{\widehat{h}_{I_x^n}}{h_{I_x}}\stackrel{p}{\longrightarrow}1, & \mbox{if }a_n\rightarrow a>0\\
\frac{\widehat{h}_{I_x^n}}{h^{opt}(x)}\stackrel{p}{\longrightarrow}1, & \mbox{if } a_n\rightarrow 0 \mbox{ and } na_n\rightarrow\infty;
\end{eqnarray*}

\item if, in addition, $Q_2(n)\rightarrow 0$ for some
$\lambda>0$, then
\begin{eqnarray*}
\frac{\widehat{h}_{I_x^n}}{h_{I_x}}\stackrel{a.s.}{\longrightarrow}1, & \mbox{if }a_n\rightarrow a>0\\
\frac{\widehat{h}_{I_x^n}}{h^{opt}(x)}\stackrel{a.s.}{\longrightarrow}1, & \mbox{if } a_n\rightarrow 0 \mbox{ and } na_n\rightarrow\infty.
\end{eqnarray*}
\end{itemize}
where $Q_1(n)$ and $Q_2(n)$ are defined in (\ref{Qs}).
\end{theorem}
By Theorem \ref{theorem2} we get the global bandwidth estimator if we set the parameter $a$ large enough with respect to the support of the volatility function. It is easy to apply Theorem \ref{theorem2} to have the consistency of the estimator for the volatility function. So, we can state the following Corollary from Theorem \ref{theorem2}. The proof is straightforward by applying Theorem \ref{theorem2} and Theorem 3.1 of \cite{HarTsy97}. 
\begin{corollary}\label{rem2} Suppose that assumptions (a1) -- (a6) and (b1) -- (b3) hold. If  we consider the estimated bandwidth, say $\hat h$, both for the cases of global and local, it follows that
\begin{equation}
\label{rate_s}
\left|\widehat{ \sigma}^2(x;\hat h) -\sigma^2(x)\right|=O_p(n^{-2/5}).
\end{equation}
\end{corollary}
\begin{remark} \label{rem2bis}
Theorem \ref{theorem2}, by Propositions \ref{proposition1} and \ref{proposition2} in the Appendix, uses the Neural Networks estimator for the functionals $\mathbb{B}_{w_{I_x}}$ and $\mathbb{V}_{w_{I_x}}$ to overcome the issue of the pilot bandwidth estimation. We have both the consistency and optimality for the global and local bandwidth estimators. It is clear that Theorem \ref{theorem2} holds again if we consider \textit{any} consistent estimator of the functionals $\mathbb{B}_{w_{I_x}}$ and $\mathbb{V}_{w_{I_x}}$, not necessarily based on the Neural Network technique.
\end{remark}  

\section{Nonparametric confidence intervals for volatility}
\label{Conf}
Using the GAS procedure, we can build \emph{unbiased} confidence intervals for the volatility
function. An application to real data is
reported in Section \ref{data}.

The bias of the volatility estimator is given in (\ref{fun1}) and it can be estimated by (\ref{Rhat}). 
 
Without loss of  generality, for a given $a>0$ and $I_x=[x-a/2,x+a/2]$, suppose that $\mathbb{B}_{\omega_{I_x}}\approx \mathcal{B}(x)$ and $\mathbb{V}_{\omega_{I_x}}\approx \mathcal{V}(x)$, where $\mathcal{B}(x)$ and $\mathcal{V}(x)$ are defined in (\ref{fun2}). We can state the following result.
\begin{theorem}\label{corol}
Suppose that the assumptions (a1) -- (a6) and (b1) -- (b3) hold.  If  $Q_1(n)\rightarrow 0$ as $n\rightarrow\infty$, then for each $x$
\[
\sqrt{n\widehat{ h}_{I_x}}\frac{\left[\widehat{\sigma}^2(x;\widehat{ h}_{I_x})-\sigma^2(x)-\frac{1}{2}\widehat{ h}^2_{I_x}\widehat{\mathbb{B}}_{\omega_{I_x}}\right]}{\left(\widehat{\mathbb{V}}_{\omega_{I_x}}\right)^{1/2}}\stackrel{d}{\longrightarrow} N(0,1).
\]
Here $\widehat{\sigma}^2(x;\widehat{ h}_{I_x})$ is the LLE for the volatility function given in (\ref{phi.hat}). The estimators $\widehat{\mathbb{B}}_{\omega_{I_x}}$ and $\widehat{\mathbb{V}}_{\omega_{I_x}}$ are defined in (\ref{Rhat}). The estimated optimal bandwidth $\widehat{ h}_{I_x}$ is given in section \ref{idea}. $Q_1(n)$ is defined in (\ref{Qs}). 
\end{theorem}

If we drop the assumptions $\mathbb{B}_{\omega_{I_x}}\approx \mathcal{B}(x)$ and $\mathbb{V}_{\omega_{I_x}}\approx \mathcal{V}(x)$, Theorem \ref{corol} holds again but replacing $\widehat{\sigma}^2(x;\widehat{ h}_{I_x})$ and $\sigma^2(x)$ with $\widehat{\sigma}^2(I_x)$ and $\sigma^2(I_x)$, respectively, where 
\begin{equation}
\label{new.stat}
\widehat{\sigma}^2(I_x)=\frac{\sum_{t=2}^{n}\widehat{\sigma}^2(X_{t-1};\widehat{ h}_{I_x})\mathbb{I}(X_{t-1}\in
I_x)}{\sum_{t=2}^{n}\mathbb{I}(X_{t-1}\in I_x)},\quad \sigma^2(I_x)=\frac{1}{\mu_X(I_x)}\int_{I_x}\sigma^2(x)f_X(x)dx.
\end{equation}

\section{Nonparametric testing for symmetry of volatility}
\label{Test}
Another useful application of our results in Section \ref{idea} is to build a statistical test for the symmetry of the volatility function around zero. The hypothesis $H_0$ is $\sigma^2(x)\equiv\sigma^2(-x)$ for each $x$, and the alternative $H_1$ means that $\sigma^2(x')\not =  \sigma^2(-x')$ for at least one $x'$. Without loss of generality, suppose that $\mathbb{V}_{\omega_{I_x}}\approx \mathcal{V}(x)$ for a given $a>0$ and $I_x=[x-a/2,x+a/2]$, where $\mathcal{V}(x)$ is defined as in section \ref{Conf}. We have the following result.
\begin{theorem}\label{cor2}
Assume that: a) the bivariate density function for the process $\{X_t\}$ in model (\ref{eqn01}) is bounded, say $f_{X_1X_2}(x_1,x_2)\le C_0<\infty$, $\forall (x_1,x_2)\in \mathbb{R}^2$; b) the same assumptions as in Theorem \ref{corol} hold. If  $Q_1(n)\rightarrow 0$ as $n\rightarrow\infty$, then under $H_0$ we have
\[
\sqrt{n}\frac{\left[\widehat{ h}_{I_x}^{1/2}\widehat{\sigma}^2(x;\widehat{ h}_{I_x})-\widehat{ h}_{I_{-x}}^{1/2}\widehat{\sigma}^2(-x;\widehat{ h}_{I_x})\right]}{\left(\widehat{\mathbb{V}}_{\omega_{I_x}}+\widehat{\mathbb{V}}_{\omega_{I_{-x}}}\right)^{1/2}}\stackrel{d}{\longrightarrow} N(0,1)\qquad \mbox{for\ each\ } x>0,
\]
where $\widehat{\sigma}^2(x;\widehat{ h}_{I_x})$ is the LLE for the volatility function given in (\ref{phi.hat}). The estimated optimal bandwidth $\widehat{ h}_{I_x}$ is given in section \ref{idea} and the estimator $\widehat{\mathbb{V}}_{\omega_{I_x}}$ is defined in (\ref{Rhat}). $Q_1(n)$ is defined in (\ref{Qs}). 
\end{theorem}

Now, we have to consider a number of points, say $n_x$, such that $\{-x_i,x_i\}$, $i=1,\ldots,n_x/2$. We have to do $n_x/2$ tests by Theorem \ref{cor2}. Using a simple multiple test approach as the Bonferroni's technique, we have to compute
\[
T_i=\sqrt{n}\frac{\left[\widehat{ h}_{I_{x_i}}^{1/2}\widehat{\sigma}^2(x_i;\widehat{ h}_{I_{x_i}})-\widehat{ h}_{I_{-x_{i}}}^{1/2}\widehat{\sigma}^2(-x_i;\widehat{ h}_{I_{-x_i}})\right]}{\left(\widehat{\mathbb{V}}_{\omega_{I_{x_i}}}+\widehat{\mathbb{V}}_{\omega_{I_{-x_{i}}}}\right)^{1/2}}\qquad i=1,\ldots,n_x/2.
\]
Given a level $\alpha$ as the first type error, we accept the Null if all of the following conditions are satisfied,
\[
|T_i|<q_\phi(1-\alpha/n_x)\qquad i=1,\ldots,n_x/2,
\]
where $q_\phi(\cdot)$ is the quantile from the Standard Normal distribution. In this way, we reject $H_0$ if at least one condition above is not true.

Note that the results in Theorems \ref{corol} and \ref{cor2} hold again if we drop the assumption that $\mathbb{V}_{\omega_{I_x}}\approx \mathbb{V}(x)$ and replace $\widehat{\sigma}^2(\cdot)$ with $\widehat{\sigma}^2(I_{\cdot})$, defined in (\ref{new.stat}).
 
\section{Simulation study}
\label{sim}
In the first part of the simulation study, we compare the nonparametric GAS method for volatility estimations with the classic parametric estimation methods (maximum likelihood estimator, MLE). 
It must be remarked that a direct comparison between parametric methods (MLE) and nonparametric methods (GAS) should not be made, for several reasons: nonparametric methods take advantage from being model free whereas parametric methods take advantage from having a faster convergence rate under the assumption of correct specification of the model. It is expected, therefore, to see more robust estimations from nonparametric methods and more efficient estimations from parametric methods. Therefore, they are not directly comparable since they works on different assumptions. Anyway, the following results show very interesting performances of the two estimation methods that are worthwhile to be reported. In particular, surprisingly, GAS shows (more robust results, as expected, but also) lower variability with respect to MLE for small sample sizes, even in the case of correct specification of the model.

We consider three models with a null conditional mean function. They are reported in the following table, where $\phi(\cdot)$ denotes the standard normal density.
\begin{center}
\begin{tabular}{lll}
\hline\noalign{\smallskip} 
Model &  & Errors \\ \hline
\emph{1: ARCH(1)} & $X_t=\sqrt{0.1+0.5X_{t-1}^2}\varepsilon_{t}$ & $\varepsilon_t\sim \phi$\\
\emph{2: GARCH(1,1)} & $X_t=\sqrt{0.1+0.3X_{t-1}^2+0.2\sigma^2_{t-1}}\varepsilon_{t}$ & $\varepsilon_t\sim \phi$ \\
\emph{3: HT} & $X_t=\left[\phi(X_{t-1}+1.2)+1.5\phi(X_{t-1}-1.2)\right]\varepsilon_{t}$
& $\varepsilon_t\sim \phi$\\
\hline
\end{tabular}
\end{center}

Model 1 is a classic $ARCH(1)$ and model 2 a $GARCH(1,1)$. Model 3 (HT) is a nonlinear $ARCH$ used by \cite{HarTsy97}. All the models satisfy the assumptions of this paper. Contrary to models 1 and 2, model 3 is not symmetric with respect to zero and it is highly nonlinear, so that a variable bandwidth should be preferred for this model. We stress here that the GAS method automatically detects the kind of bandwidth to use, which is a trade-off between local and global smoothing, by automatically setting the optimal value for the parameter $a$. Anyway, given the aims of this paper, here we do not investigate on the performance of GAS with respect to the selection of the parameter $a$ (see \cite{GioPar14} for some results on this). So, for the sake of comparison, in the whole simulation study we will impose a global bandwidth for all three models.

We use R to perform a Monte Carlo simulation study with $500$ replications and three different lengths for the simulated time series: $n=(500, 1000, 2000)$. We implement the procedure described in section \ref{idea}, using the Epanechnikov kernel $K(\cdot)$ for the LLE and the Logistic Sigmoidal function for the Neural Networks estimator. The number of nodes in the hidden layer of the neural network is selected following an automatic BIC optimization procedure, as in \cite{FarCha98}. Some experiments not reported in this paper show that the number of nodes of the neural network does not have a strong influence on the final estimation results. See \cite{GioPar14} for further details.

For each replication, the integrated squared error ($ISE$) is calculated as
\begin{equation}\label{ISE}
ISE(\widehat{ h})=\frac{1}{n_x}\sum_{j=1}^{n_x}\left[\widehat{ \sigma}^2(x_j)-\sigma^2(x_j)\right]^2,
\end{equation}
where $x_1,\ldots,x_{n_x}$, with $n_x=20$, are randomly chosen over the support of the volatility function. In the following table,
we report the mean, the median and the standard deviation of the $ISE(\widehat{ h})$ ($MISE$, $MEDISE$, and $SDISE$, respectively) for the 500 replications of the models. We compare two kinds of estimators $\widehat{ \sigma}^2(x_j)$ in the (\ref{ISE}).
From the one hand we have the GAS volatility estimator $\widehat{ \sigma}^2(x) \equiv\widehat{ \sigma}^2(x;\widehat{ h}) $ given in (\ref{phi.hat}), with the bandwidth estimated by the procedure explained in section \ref{idea} (the parameter $a$ is set to a high value to have a global smoothing); we denote this estimator with the suffix \emph{GAS} in the tables. From the other hand, we use the classic MLE for the estimation of the parameters of a parametric $GARCH(1,1)$ model, using the package \texttt{rugarch} of R.

\begin{table}[t]
\centering
\begin{tabular}{rcccccc} \hline
\multicolumn{7}{l}{\textbf{Model 1: $ARCH(1)$}} \\ \hline
$n$ &       $ MISE_{GAS} $ &  $MISE_{MLE}$ & $MEDISE_{GAS}$ & $MEDISE_{MLE}$  & $SDISE_{GAS}$ &  $SDISE_{MLE}$ \\ \hline
 500 & 0.563642 & 0.315009 & 0.324258 & 0.187120 & 1.223880 & 0.442006 \\
1000 & 0.490322 & 0.391450 & 0.295758 & 0.248785 & 0.549949 & 0.491264 \\
2000 & 0.520154 & 0.549357 & 0.380937 & 0.360051 & 0.606490 & 0.658968 \\
\hline
 \multicolumn{7}{l}{\textbf{Model 2: $GARCH(1,1)$}} \\ \hline
$n$ &       $ MISE_{GAS} $ &  $MISE_{MLE}$ & $MEDISE_{GAS}$ & $MEDISE_{MLE}$  & $SDISE_{GAS}$ &  $SDISE_{MLE}$ \\ \hline
 500 & 0.582419 & 1.54658 & 0.409780 & 0.608519 & 0.546963 & 2.99049 \\
1000 & 0.716475 & 1.37249 & 0.515701 & 0.657613 & 0.690708 & 2.16681 \\
2000 &  1.071458 & 1.30674 & 0.775978 & 0.804068 & 1.127064 & 1.97655 \\
\hline
\multicolumn{7}{l}{\textbf{Model 3: $HT$}} \\ \hline
$n$ &       $ MISE_{GAS} $ &  $MISE_{MLE}$ & $MEDISE_{GAS}$ & $MEDISE_{MLE}$  & $SDISE_{GAS}$ &  $SDISE_{MLE}$ \\ \hline
 500 &  0.364365 & 0.957462 & 0.255328 & 0.907852 & 0.356133 & 0.337183\\
1000 & 0.356558 & 1.732367 & 0.261877 & 1.661635 & 0.331395 & 0.496300\\
2000 & 0.385294 & 3.403906 & 0.282595 & 3.371010 & 0.331421 & 0.920338\\
\hline
\end{tabular}
\caption{\label{tabella1} \small Comparison between GAS and MLE methods for the estimation of the volatility function. The table reports the mean, the median and the standard deviation of the integrated squared error ($MISE$, $MEDISE$ and $SDISE$,
respectively) for the 500 replications of models 1-3. All the values have been multiplied by $n$ to make them more comparable.}
\end{table}
Table \ref{tabella1} shows the results. All the values of $MISE$, $MEDISE$ and $SDISE$ have been multiplied by the sample size $n$, to make them more comparable (this explains why the values shown in the table do not decrease with $n$, actually they do, after dividing by $n$). As expected, for model 1, that is an $ARCH(1)$, the smaller values are observed for the MLE method. In fact, for this model the rate of convergence of the MLE estimator is $O_p(n^{-1/2})$, faster than the convergence rate of the GAS estimator, which is $O_p(n^{-2/5})$ (see Corollary \ref{rem2}). However, for $n=2000$ the results do not present any relevant difference. Instead, for model 3, we observe the smaller values for the $GAS$ method, as we expect since in this case the MLE works with a misspecified model. For this model, the GAS estimator is consistent whereas the MLE for $GARCH(1,1)$ is not. In fact, the $MISE$, $MEDISE$ and $SDISE$ (multiplied by $n$) seem to be constant for GAS when $n$ grows. This is not true for the MLE, where instead they increase. 

But very surprisingly, for model 2 which is a $GARCH(1,1)$, we again observe the smaller values for the GAS method notwithstanding the MLE works with a correctly specified model. This result actually gives evidence of the usefulness of Theorems \ref{theorem1} and \ref{theorem2}. In fact, thanks to Theorem \ref{theorem1}, the GAS method formulates and estimates the $GARCH(1,1)$ by means of a particular $NARCH(1)$, basing on a unique regressor $X_{t-1}$, thus with rate $O_p(n^{-2/5})$. Moreover, by Theorem \ref{theorem2} we estimate the optimal bandwidth. On the other side, the MLE works with a (correctly specified) model with two regressors ($X_{t-1}$ and $\sigma_{t-1}$), one of which is latent and therefore estimated. As a consequence, its finite sample performance shows a penalty for this aspect. However, when $n$ grows the differences tend to reduce.

Finally, we report some simulation results to evaluate the test for the symmetry of the volatility function, proposed in section \ref{Test}. We have applied the test for model 2, where the volatility function is symmetric, and for model 3, where the volatility function is not symmetric (\emph{leverage effects}). 
%We do not consider model 2, the $GARCH(1,1)$, because for this model the test cannot be made without transforming the $GARCH$ by a $NARCH$, as model 3.
We consider 500 simulation runs and $n_x=20$ points (equally spaced) with $10$ multiple tests. We use the global bandwidth in $T_i$, $i=1,\ldots,10$ and the estimator $\widehat{\mathbb{V}}_{\omega_{I_x}}$ in (\ref{Rhat}). We set $\alpha=1\%$.
The results in table \ref{tabella4} have to be read as the size of the test for model 2, which is symmetric, and the power of the test for model 3, which is asymmetric. The first row is around the nominal size of $1\%$ for large $n$. Moreover, as we expect, the power (second row) grows when $n$ increases.  
\begin{table}
\centering
\begin{tabular}{lccc} \hline
& & $n$ &\\
Model &  500 & 1000 & 2000\\ \hline
$GARCH(1,1)$ &  3.8\% & 1.6\% & 1.6\%\\
$HT$ & 70.0\% &  93.8\% & 99.8\%
\\ \hline
\end{tabular}
\caption{\label{tabella4} Empirical percentages to reject the Null (symmetry) over 500
replications from models 2 ($GARCH(1,1)$) and 3 ($HT$), with $n={500, 1000, 2000}$. $\alpha=1\%$.}
\end{table}

\section{Real data Application}
\label{data}
\begin{figure}[h]
    \centering
 \resizebox{10cm}{!}{\includegraphics{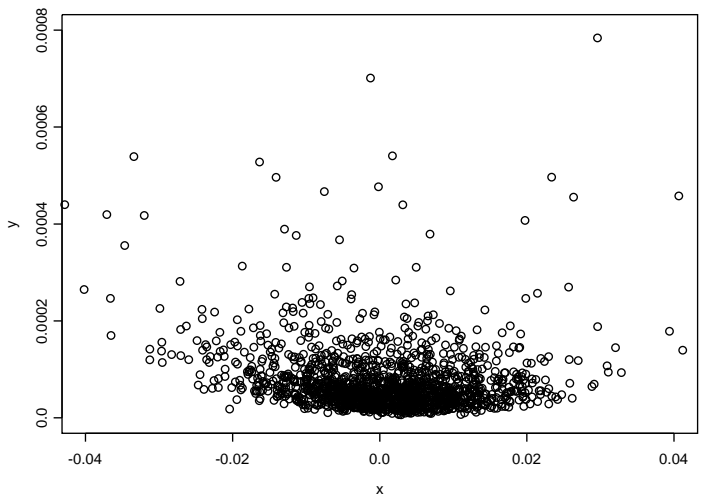}}
    \caption{The observed returns of Dow Jones index from 1996, January 3$^{rd}$ to the end of January 2002 are on the "x" axis. The "Realized Volatility" values are on "y" axis}
    \label{fig1}
\end{figure}

In this section we apply our method to real data. We consider a
time series of Dow Jones index from 1996, January 3$^{rd}$ to the
end of January 2002. It means that the length of time series is 1500. We derive the returns and use them in order to estimate the volatility function and its confidence intervals using the GAS procedure.

As a proxy of the true volatility, we also extract the \emph{realized volatility}
time series from the Oxford-Man Institute's \emph{realized library}, which contains daily measures of how volatility financial assets or indexes were in the past, basing on infra-daily data (see \cite{HebAlt09}).
In figure \ref{fig1}, we report the returns on the $x$ axis and the realized volatility on the $y$ axis. 

\begin{figure}[t]
    \centering
 \resizebox{10cm}{!}{\includegraphics{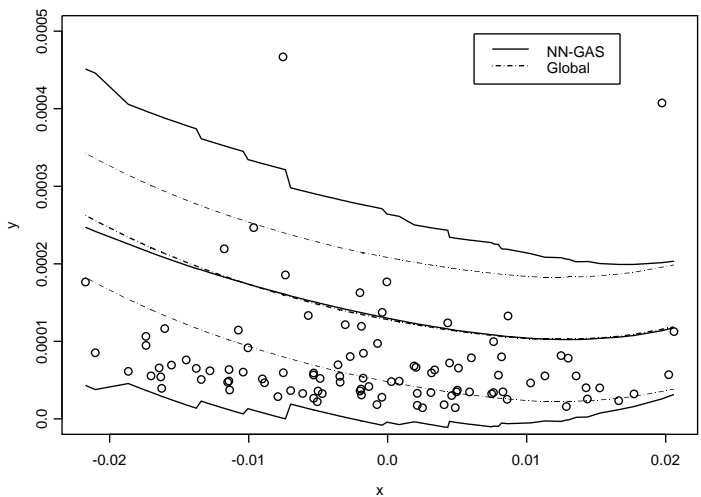}}
    \caption{The central solid line is the volatility function estimated by GAS method. The solid lines on the top and bottom are the upper and lower confidence intervals at 95\%, respectively. The dashed lines refer to the estimated volatility and confidence intervals using the global bandwidth. The dot points are the realized volatility values.}
    \label{fig2}
\end{figure}

Using only the observed returns, we apply our method to estimate the volatility function. We draw it in figure \ref{fig2} as the central solid line. The estimate of the parameter $a$ (for bandwidth slection) is $\widehat{a}=0.089$, following \cite{GioPar14}. 
By Theorem \ref{corol}, we can build the confidence intervals. They are shown in figure \ref{fig2} by the two external solid lines. We add the estimated volatility function and the confidence intervals derived by imposing a \emph{global constant smoothing} (i.e., with constant bandwidth obtained by fixing a large $a$). They are shown in figure \ref{fig2} by the central dashed and the two external dashed lines, respectively. Note that, in both cases of local and global approaches, we do not consider any correction for the bias.

We can point out an important difference between the GAS and the global bandwidth approaches. If we look at the confidence intervals in figure \ref{fig2}, the GAS ones have a better adaptability than the global bandwidth method. In fact, the GAS procedure
has the advantage to take into account the heteroscedastic behaviour in the data. In figure (\ref{fig2}) we plot $n_x=100$ points (returns and realized volatility) which are randomly chosen. By figures (\ref{fig1}) and (\ref{fig2}) we can note that there is an asymmetry for the realized volatility. In particular, we can observe a greater variability for negative values of returns. All
that is confirmed by the GAS confidence intervals (solid lines in figure (\ref{fig2})) which are wider than the confidence intervals for the global bandwidth approach (dashed lines) when the observed returns are negative. Finally, for the $n_x=100$ points in figure (\ref{fig2}), we have an actual coverage of 98\% and 45\% for the GAS and global bandwidth methods, respectively. It means that we have an
important gain with respect to the global bandwidth technique when we need to consider an asymmetric behaviour in the data.

%%%%%%%%%%%%%%%%%%%%%%%%%%%%%%%%%%%%%%%%%%%%%%%%%%%%%%%%%%%%%%%%%%%%%%%%%%%%%%%%%%%%%%%%%%%%%%%%%%%%%%%%%%%%%%%%%%%%%%%%%%%%
\section{Conclusions}
\label{final}
In this paper we have presented a general nonparametric framework for volatility analysis. The main contributions of this work are:
\begin{itemize}
\item the extension of the GAS method of \cite{GioPar14} to the framework of dependent data, to achieve an optimal bandwidth estimation for volatility;
\item a new nonparametric estimator of the volatility function is proposed, based on local linear polynomials with data-driven optimal local bandwidth. The new volatility estimator reaches the optimal convergence rate, as shown theoretically in the paper;
\item moreover, starting from the functionals that we need to estimate for the optimal bandwidth in GAS procedure, we can use them to derive two useful inferential tools to test the validity of a given parametric model: nonparametric confidence intervals and test for symmetry;
\item last, but not least, a new representation of the $GARCH(1,1)$ model by means of a nonparametric $ARCH(1)$ model. With this new representation, we avoid the use of the (latent) lagged volatility in the model and, therefore, a more precise \emph{News Impact Curve} can be derived and estimated. Moreover, we improve the rate of convergence of the nonparametric volatility estimator.
\end{itemize}

\appendix
\section{Assumptions and Proofs}
We make the following assumptions. First, given (\ref{eqn01}) and (\ref{mphi.x}), we need to guarantee that $E(X_{t+1})^4<\infty$.

\vspace{10pt}\noindent\textbf{Assumptions (a)}
\begin{enumerate}
\item[(a1)] The errors $\varepsilon_t$ have a continuous and positive density function with
\[
E(\varepsilon^2_t)=1,\quad E(\varepsilon_t)=E(\varepsilon^3_t)=0, \quad E(\varepsilon_t)^4<\infty.
\]
\item[(a2)] The function $\sigma(\cdot)$ is positive and has a continuous second derivative.
\item[(a3)] There exist some constants $M$ and $\alpha$ such that,
\begin{itemize}
\item[i)] $0<M<\infty$, $\sigma(y) \leq M(1+|y|)$ and $M\left[E|\varepsilon_t|^4\right]^{1/4}<1$ for all $y\in{\mathbb{R}}$;
\item[ii)] $0\leq\alpha\leq M$, $\sigma(y)-\alpha|y|=o(1)$ for $|y|\rightarrow\infty$.
\end{itemize}
\item[(a4)] The process $\{X_t\}$ is strictly stationary.
\item[(a5)] The density function $f_X(\cdot)$ of the (stationary) measure of the process $\mu_X$ exists; it is bounded, continuous and positive on every compact set in ${\mathbb{R}}$.
\item [(a6)] The kernel function, $K(\cdot)$, is compactly supported bounded function such that it is positive on a set of positive Lesbegues measure.
\end{enumerate}

\vspace{5pt}\noindent Under the
assumptions (a1), (a3), (a5) and (a2) with the part that the function $\sigma(\cdot)$ is always positive, it can be shown that the process is
geometrically ergodic and exponentially $\alpha$-mixing with $E(X_t^4)<\infty$ (see
\cite{HarTsy97}). Moreover, assumption (a4) is only made to simplify the proofs. Assumption (a2), with the part of the continuous second derivative for $\sigma(\cdot)$, is used for the estimation of the same second derivative for $\sigma^2(\cdot)$. In particular, we need that this second derivative is continuous in order to apply the bounds for the Neural Networks estimation. Finally, assumption (a6) is typical for the Kernel function as in \cite{HarTsy97}.

\vspace{10pt}\noindent\textbf{Assumptions (b)}
\begin{enumerate}
\item[(b1)] $\sum_{k=1}^{d_n}\left|c_k\right|\le \Delta_n$.
\item[(b2)] $d_n\rightarrow \infty$, $\Delta_n\rightarrow \infty$ as
$n\rightarrow \infty$.
\item[(b3)] The activation function is strictly increasing, sigmoidal and has a continuous second derivative.
\end{enumerate}

\vspace{5pt}\noindent The assumptions (b1) and (b2) for $d_n$ are typical in order to assure the approximation capability of the Neural Networks technique. Instead, the assumptions (b1) and (b2) for $\Delta_n$ allow that the approximation capability of the Neural Networks works well in a non-compact sets (see \cite{FraDia06}). The assumption (b3) assures that the activation function of the Neural Networks is regular enough (continuous second derivative) in order to estimate some functionals which depend on the second derivative of  the unknown volatility function.   

\vspace{10pt}\noindent\textbf{Proof of Theorem \ref{theorem1}: }Suppose that the distribution function of $\varepsilon_t$ is symmetric around zero. Then
$\sigma_t^2=\alpha_0+\alpha_1\varepsilon_{t-1}^2\sigma^2_{t-1}+\beta \sigma_{t-1}^2=\alpha_0+(\alpha_1+\beta)\widetilde{X}_{t-1}^2$,
where $\widetilde{X}_{t}^2=\widetilde{\varepsilon}^2_{t}\sigma_t^2$ with
\[
\widetilde{\varepsilon}_t=sgn(\varepsilon_t)\sqrt{\frac{\varepsilon_t^2+\beta/\alpha_1}{1+\beta/\alpha_1}}\qquad \mbox{ and }\qquad sgn(x)=\left\{\begin{array}{ll}1&\mbox{ if }x>0\\
0&\mbox{ if }x=0.\\
-1&\mbox{ if }x<0
\end{array}
\right.
\]
Therefore, we can write the \textit{GARCH(1,1)} process as
\begin{equation}\label{garch2}
X_t=\widetilde{X}_tC_{\widetilde{\varepsilon}_t}^{1/2}\qquad \mbox{with} \qquad C_{\widetilde{\varepsilon}_t}=1+\beta/\alpha_1\left(1-1/\widetilde{\varepsilon}_t^2\right).
\end{equation}
Note that $\widetilde{X}_t\sim ARCH(1;\alpha_0,\alpha_1+\beta)$ with the error term $\{\widetilde{\varepsilon}_t\}$ defined above. Moreover, model (\ref{garch2}) is an exact representation for the \textit{GARCH(1,1)}.

It is easy to verify that the \textit{GARCH(1,1)} model in (\ref{garch2}) is well defined in the sense that $E\left(\widetilde{\varepsilon}_tC_{\widetilde{\varepsilon}_t}^{1/2}\right)=0$ and $E\left(\widetilde{\varepsilon}_t^2C_{\widetilde{\varepsilon}_t}\right)=1$. Moreover, it follows that $E\left(X_t^2|\widetilde{X}_{t-1}=\widetilde x\right)=\alpha_0+(\alpha_1+\beta)\widetilde x^2\equiv \sigma^2(\widetilde x)$.
Now,  we can write $X_t^2=C_{\widetilde{\varepsilon}_t}\widetilde{\varepsilon}_t^2\left(\alpha_0+(\alpha_1+\beta)\frac{X_{t-1}^2}{C_{\widetilde{\varepsilon}_{t-1}}}\right)$.
So we have
\[
E\left(X_t^2|X_{t-1}=x\right)=\alpha_0+(\alpha_1+\beta)x^2E\left(\left.\frac{1}{C_{\widetilde{\varepsilon}_{t-1}}}\right|X_{t-1}=x\right),\quad \forall x\not= 0.
\]
Let  $g(x;\alpha_1,\beta)\equiv E\left(\left.\frac{1}{C_{\widetilde{\varepsilon}_{t-1}}}\right|X_{t-1}=x\right)$, $\forall x\not= 0$. When $x=0$ we can use (\ref{garch}). So we can write
\[
E(X_t^2|X_{t-1}=0)=\alpha_0+\beta E(\sigma_{t-1}^2|X_{t-1}=0)=\alpha_0+\beta E(\sigma_{t-1}^2|\varepsilon_{t-1}=0)=
\]
\[
=\alpha_0+\beta E(\sigma_{t-1}^2)=\alpha_0+\frac{\beta \alpha_0}{1-\alpha_1-\beta}\equiv A_0.
\]
Let $B_0=\beta\alpha_0/(1-\alpha_1-\beta)$.
Now, we need to evaluate the function $g(x;\alpha_1,\beta)$, for each $x\not =0$. In such a case, we have $X_{t-1}\not =0 \Longleftrightarrow\varepsilon_{t-1}\not =0\Longleftrightarrow C_{\widetilde{\varepsilon}_{t-1}}>0$, with probability one. Since $C_{\widetilde{\varepsilon}_{t-1}}<\infty$, with probability one, the function $g(\cdot;\alpha_1,\beta)$ is always positive and bounded for each $x\not =0$. Now, we can conclude that
\begin{eqnarray}\label{garch11}
E(X_t^2|X_{t-1}=x) &=&\left\{\begin{array}{lll}
A_0&\mbox{if}&x=0,\\
A_0+(\alpha_1+\beta)\tilde g(x;\alpha_1,\beta)x^2&\mbox{if}&x\not =0, 
\end{array}
\right.
\end{eqnarray}
where $\tilde g(x;\alpha_1,\beta)=g(x;\alpha_1,\beta)-\frac{B_0}{x^2(\alpha_1+\beta)}$.

Finally, if $|x|\rightarrow\infty$ then $|X_{t-1}|\ge|x|$,with probability one, since $C_{\widetilde{\varepsilon}_{t-1}}$ is always bounded. Thus we have $X_{t-1}=O(\widetilde{X}_{t-1})$, with probability one. Moreover, $E(X_t^2)=E(\widetilde{X}_t^2)$. Then, there exists a $M>0$ such that 
\[
E(X_t^2|X_{t-1}=x)=E(X_t^2|\widetilde{X}_{t-1}=x)=\alpha_0+(\alpha_1+\beta)x^2\qquad \forall x>M.
\] 
It follows that $g(x;\alpha_1,\beta)\rightarrow 1$ and also $\tilde g(x;\alpha_1,\beta)\rightarrow 1$ when $|x|\rightarrow\infty$.\hfill $\Box$

\vspace{10pt}\noindent\textbf{Proof of Theorem \ref{theorem2}: } We
analyze the convergence in probability since the almost sure convergence
is straightforward as in the proofs of Propositions \ref{proposition1} and \ref{proposition2} in section \ref{GioPar16}.

First, consider $I_x$. By the assumptions of this Theorem, we have that
$\widehat{\mathbb{B}}_{\omega_{I_x}}>0$ in probability, if
$n\rightarrow\infty$. Therefore, by Propositions \ref{proposition1} and \ref{proposition2} in section \ref{GioPar16} we have that
$\widehat{h}_{I_x}/h_{I_x}\stackrel{p}{\longrightarrow}1$, when $n\rightarrow \infty$. Since $I_x^n\rightarrow I_x$ when $a_n\rightarrow a$ and given that $h_{I_x}$ is a bounded and continuous function with respect to $a$, it follows that $\widehat{h}_{I_x^n}/h_{I_x}\stackrel{p}{\longrightarrow}1$ when $a_n\rightarrow a$ with $n\rightarrow\infty$.

Now we can consider the case when $a_n\rightarrow 0$. Using the
mean value theorem it follows that
$h_{I_x^n}/h^{opt}(x)\rightarrow 1$, when $n\rightarrow \infty$.
So we have only to prove that
\begin{displaymath}
\frac{\widehat{h}_{I_x^n}}{h_{I_x^n}}\stackrel{p}{\longrightarrow}1
\quad n\rightarrow \infty
\end{displaymath}
It is sufficient to show that the number of values from the
process (\ref{eqn01}) in $I_x^n$, $N_{I^n_x}$, tends to infinity
with probability one, if $n\rightarrow\infty$, in order to
apply, again, Propositions \ref{proposition1} and \ref{proposition2} in section \ref{GioPar16}.

We fix a positive $a'$ in the sequence $\{a_n\}$. Thus, we have a
$I_x^{a'}$ and $J_x^{a'}:=\overline{I}_x^{a'}$. We can build a Markov Chain
with two states, $I$ and $J$, which are the states when the
process from (\ref{eqn01}) is in $I_x^{a'}$ and $J_x^{a'}$, respectively.
Let $p_{JI}^{(n)}$ be the transition probability from state $J$ to
state $I$ in $n$ steps. Now, using $\mu_X(\cdot)$ we get the
unique stationary probability. Based on Markov's Theorem it follows that
$p_{\cdot I}^{(n)}\rightarrow \mu_X(I_x^{a'})$, when $n\rightarrow
\infty$, for every initial state $J$.

Based on assumptions (a) the process in (\ref{eqn01}) is
geometrically ergodic, so there exists a $n_0$ such that $\forall
n>n_0$ and $\forall a'>0$ we have $\left|p_{\cdot
I}^{(n)}-\mu_X(I_x^{a'})\right|\le W_1 e^{-W_2n}$, with $W_1$ and
$W_2$ two positive constants.\\ Now, using the Ergodic Theorem for
Markov's Chain we can write $N_{I^{a'}_x}\approx n p_{\cdot I}^{(n)}$
with probability one. By assumption (a5), it is
$\mu_X(I_x^{a'})=f_X(x')a'$, for a value $x'\in I_x^{a'}$. Therefore, $n p_{\cdot
I}^{(n)}\approx W_1ne^{-W_2n}+f_X(x')na'$. Since
$na_n\rightarrow\infty$, when $n\rightarrow\infty$, if we replace
$a'$ with $a_n$, then we have that $N_{I_x^n}\rightarrow\infty$
with probability one. Finally, the result follows. \hfill $\Box$\\

\begin{remark} \label{rem3} looking at the proof of
Theorem \ref{theorem2}, we can say that the condition $na_n\rightarrow\infty$
can be replaced by the assumption that the number of values from
the process (\ref{eqn01}) in $I_x^n$ must tend to infinity but
with a lower order with respect to $n$, when $n\rightarrow\infty$.
\end{remark}

\noindent{\textbf{Proof of Theorem \ref{corol}}}:
By Theorem \ref{theorem2} we have that $\widehat{ h}_{I_x}/h_{I_x}\stackrel{p}{\longrightarrow} 1$. Using the same arguments as in Proposition \ref{proposition1} in section \ref{GioPar16}, it follows that $\widehat{\mathbb{B}}_{\omega_{I_x}}\stackrel{p}{\longrightarrow}\mathbb{B}_{\omega_{I_x}}\approx \mathcal{B}(x)$. Besides, by Proposition \ref{proposition2} in section \ref{GioPar16} we have that  $\widehat{\mathbb{V}}_{\omega_{I_x}}\stackrel{p}{\longrightarrow}\mathbb{V}_{\omega_{I_x}}\approx \mathcal{V}(x)$. The quantities $\mathcal{B}(x)$ and $\mathcal{V}(x)$ are defined in (\ref{fun2}). Moreover, by  \ref{rate_s} in Remark \ref{rem2}, $\widehat{\sigma}^2(x;\widehat{ h}_{I_x})-\widehat{\sigma}^2(x;h_{I_x})\stackrel{p}{\longrightarrow}0$. Therefore, we can conclude that
\[
\sqrt{n\widehat{ h}_{I_x}}\frac{\left[\widehat{\sigma}^2(x;\widehat{h}_{I_x})-\sigma^2(x)-\frac{1}{2}\widehat{ h}^2_{I_x}\widehat{\mathbb{B}}_{\omega_{I_x}}\right]}{\left(\widehat{\mathbb{V}}_{\omega_{I_x}}\right)^{1/2}}\quad \mbox{ and }\quad \sqrt{n h_{I_x}}\frac{\left[\widehat{\sigma}^2(x;h_{I_x})-\sigma^2(x)-\frac{1}{2}h^2_{I_x}\mathcal{B}(x)\right]}{\left(\mathcal{V}(x)\right)^{1/2}}
\]
have the same asymptotic distribution. Applying Theorem 3.2 of \cite{HarTsy97} the result follows. \hfill $\Box$
\\
\\
\noindent{\textbf{Proof of Theorem \ref{cor2}}}:
Under $H_0$, $\sigma^2(x)=\sigma^2(-x)$, $\forall x>0$. The same is true for the bias and variance.  We assume that $\mathbb{V}_{\omega_{I_x}}\approx \mathcal{V}(x)$. By Theorem \ref{theorem2} and Proposition \ref{proposition2} in section \ref{GioPar16}, we have $\widehat{ h}_{I_x}/h_{I_x}\stackrel{p}{\longrightarrow}1$ and $\widehat{\mathbb{V}}_{\omega_{I_x}} \stackrel{p}{\longrightarrow}\mathcal{V}(x)$. Let 
\[
\widehat{ T}(x)=\sqrt{n}\frac{\left[\widehat{ h}_{I_{x}}^{1/2}\widehat{\sigma}^2(x;\widehat{ h}_{I_{x}})-\widehat{ h}_{I_{-x}}^{1/2}\widehat{\sigma}^2(-x;\widehat{ h}_{I_{-x}})\right]}{\left(\widehat{\mathbb{V}}_{\omega_{I_{x}}}+\widehat{\mathbb{V}}_{\omega_{I_{-x}}}\right)^{1/2}}.
\]
It follows that $\widehat{ T}(x)$ has the same asymptotic distribution as 
\[
\widetilde{T}(x)=\sqrt{nh_{I_x}}\frac{\left[\widehat{\sigma}^2(x;h_{I_x})-\widehat{\sigma}^2(-x;h_{I_{-x}})\right]}{\left(2\mathcal{V}(x)\right)^{1/2}}.
\]
By Theorem \ref{corol}, both $\sqrt{nh_{I_x}}\widehat{\sigma}^2(x;h_{I_x})$ and $\sqrt{nh_{I_{x}}}\widehat{\sigma}^2(-x;h_{I_{-x}})$ have an asymptotic Normal distribution. We have only to evaluate the mixed terms, that is, the terms with different fixed points, $-x$ and $x$. Now, it is sufficient to prove that
\[
(I)\mbox{ } hE\left[\frac{1}{h^2}K\left(\frac{X-x}{h}\right)K\left(\frac{X+x}{h}\right)\right]\rightarrow 0 \qquad \mbox{and}
\]
\[
(II) \mbox{ } hE\left[\frac{1}{h^2}K\left(\frac{X_1-x}{h}\right)K\left(\frac{X_2+x}{h}\right)\right]\rightarrow 0
\]
when $n\rightarrow\infty$, with $h\rightarrow 0$. The univariate random variable $X$ is drawn form the process $\{X_t\}$ in model (\ref{eqn01}). While $(X_1,X_2)$ is a bivariate random variable from the same process.
\[
I=\frac{1}{h}\int K\left(\frac{X-x}{h}\right)K\left(\frac{X+x}{h}\right)f_X(X)dX=\int K(Z)K\left(Z+\frac{2x}{h}\right)f_X(x+hZ)dZ,
\]
changing the variable from $X$ to $Z=\frac{X-x}{h}$. By assumption (a5) it follows that
\[
\int K(Z)K\left(Z+\frac{2x}{h}\right)f_X(x+hZ)dZ\le \sup f_X(x+hZ)\int K(Z)K\left(Z+\frac{2x}{h}\right)dZ\rightarrow 0 
\]
when $n\rightarrow\infty$, $h\rightarrow 0$ and $K(\cdot)$ is bounded by (a6). Note that the convergence to zero holds for any rate with respect to $h$. Finally, 
\[
II=\frac{1}{h}\int\int K\left(\frac{X_1-x}{h}\right)K\left(\frac{X_2+x}{h}\right)f_{X_1X_2}(X_1,X_2)dX_1dX_2=
\]
\[
=h\int\int K(Z_1)K\left(Z_2+\frac{2x}{h}\right)f_{X_1X_2}(x+hZ_1,x+hZ_2)dZ_1dZ_2,\quad (III)
\]
changing the variable from $(X_1,X_2)$ to $(Z_1=\frac{X_1-x}{h},Z_2=\frac{X_2-x}{h})$. Since $f_{X_1X_2}(\cdot,\cdot)$ is bounded by $C_0$, then
\[
III\le hC_0\int K(Z_1)dZ_1\int K\left(Z_2+\frac{2x}{h}\right)dZ_2=hC_0\int K\left(Z_2+\frac{2x}{h}\right)dZ_2\rightarrow 0
\]
when $n\rightarrow\infty$, $h\rightarrow 0$ and again for the boundedness of $K(\cdot)$ by (a6). The proof is complete. \hfill $\Box$

\section{Supplementary results}\label{GioPar16}

In this Section we show a self-contained method to estimate the unknown functionals for the asymptotic bandwidth parameter in Local Linear Polynomial estimator. We extend the method in \cite{GioPar14} to the case of dependent data. Moreover, we use a different technical approach as in \cite{FraDia06} and \cite{Gyorfy02} in order to deal with the Neural Networks estimator for unknown functions defined on non compact sets.

\begin{remark}\label{rem1} Under the
assumptions ({a1})--({a5}), it can be shown that the process is
geometrically ergodic and exponentially $\alpha$-mixing (see
\cite{HarTsy97}).
\end{remark}

Let us consider, for some $\lambda>0$,
\begin{equation}
\label{Qs}
Q_1(n):=\frac{\Delta_n^2d_n\log\left(\Delta_n^2d_n\right)}{\sqrt{n}}, \quad
Q_2(n):=\frac{\Delta_n^4}{n^{1-\lambda}}, \quad
\mathcal{F}_n:=\left\{q:\sum_{k=1}^{d_n}\left|c_k\right|\le
\Delta_n\right\}.
\end{equation}
$\mathcal{F}_n$ is the class of feedforward neural networks with
bounded weights. Now
$\mathcal{F}=\bigcup_{n=1}^{\infty}\mathcal{F}_n$ is the class of
general feedforward neural networks. $\mathcal{F}$ is dense with
respect to the class of squared integrable functions using a
predefined measure (\cite{Hor91}). Under model
(\ref{eqn01}), the Neural Network estimator $q(x, \B{\widehat{ \eta}})$
 can be written as
\begin{equation}
\label{NNe}
q(x, \B{\widehat{ \eta}})=\arg
\min_{f\in\mathcal{F}_n}\frac{1}{n-1}\sum_{k=1}^{n-1}\left(X_{k+1}^2-f(X_k)\right)^2
\end{equation}

\subsection{Preliminary results}

In this section we report some preliminary results for the Neural Networks estimator.

Lemma \ref{lemma1} extends the results for the consistency in
\cite{FraDia06} with respect to the Neural Network estimator,
$q(x, \B{\widehat{ \eta}})$, using assumptions (a) and (b).
Moreover, the same consistency, as in \cite{FraDia06}, is shown in
the Lemma \ref{lemma2} for the Neural Network estimator of the second
derivative for the unknown function $\sigma^2(x)$.  
\begin{lemma}\label{lemma1}
Under assumptions (a1) -- (a5)
and (b), the estimator $q(x;\widehat{\eta})$ of $\sigma^2(x)$, defined
in (\ref{NNe}), is consistent in the
sense that:
\begin{itemize}
\item if $Q_1(n)\rightarrow 0$ as $n\rightarrow \infty$, then
\begin{displaymath}
E\int\left(q\left(x;\widehat{\mathbf{\eta}}\right)-\sigma^2(x)\right)^2d\mu_X(x)\rightarrow
0 \qquad n\rightarrow \infty;
\end{displaymath}
\item if, additionally, $Q_2(n)\rightarrow 0$ for some
$\lambda>0$, then \begin{displaymath}
\int\left(q\left(x;\widehat{\mathbf{\eta}}\right)-\sigma^2(x)\right)^2d\mu_X(x)\stackrel{a.s.}{\longrightarrow}
0 \qquad n\rightarrow \infty.
\end{displaymath}
\end{itemize}
\end{lemma}
\noindent\textit{\textbf{Proof:}} It is sufficient to apply
Theorem (3.2) in \cite{FraDia06} with respect to the estimator
$q(x, \B{\widehat{ \eta}})$.
Based
on the previous Remark \ref{rem1}, the process in (\ref{eqn01}) is
exponentially $\alpha$-mixing and the activation function for
Neural Network estimator is sigmoidal, continuous and strictly
increasing by (b3). So the conditions for the Theorem (3.2) in
\cite{FraDia06} are satisfied. \hfill $\Box$

\begin{lemma}\label{lemma2}
Under the same assumptions as in Lemma \ref{lemma1}, the
estimator of the second derivative of $\sigma^2(x)$ is consistent
in the sense that:
\begin{itemize}
\item if $Q_1(n)\rightarrow 0$ as $n\rightarrow \infty$, then
\begin{displaymath}
E\int\left(q^{(2)}\left(x;\widehat{\mathbf{\eta}}\right)-\sigma^2_{(2)}(x)\right)^2d\mu_X(x)\rightarrow 0 \qquad n\rightarrow \infty;
\end{displaymath}
\item if, additionally, $Q_2(n)\rightarrow 0$ for some
$\lambda>0$, then \begin{displaymath}
\int\left(q^{(2)}\left(x;\widehat{\mathbf{\eta}}\right)-\sigma^2_{(2)}(x)\right)^2d\mu_X(x)\stackrel{a.s.}{\longrightarrow} 0 \qquad n\rightarrow \infty,
\end{displaymath}
where $\sigma^2_{(2)}(x)$ is the second derivative of $\sigma^2(x)$.
\end{itemize}
\end{lemma}
\noindent\textit{\textbf{Proof:}} Define with $\mathcal{G}$ the
class of all functions $\sigma^2(x)$ satisfying the assumptions
(a2) and (a3). Now we can write \begin{displaymath}
\int\left(q^{(2)}\left(x;\widehat{\mathbf{\eta}}\right)-\sigma^2_{(2)}(x)\right)^2d\mu_X(x)\le\left\|D^2\right\|^2\int\left(q\left(x;\widehat{\mathbf{\eta}}\right)-\sigma^2(x)\right)^2d\mu_X(x)
\end{displaymath}
where $\left\|D^2\right\|^2=\sup_{f\in\mathcal{G}}\int
\left(f''(x)\right)^2d\mu_X(x)$.\\ By assumptions, the linear
operator $D^2$ is bounded. So $\left\|D^2\right\|^2<\infty$.
Finally, using Lemma \ref{lemma1} we obtain the result.\hfill $\Box$\\

\noindent The next two lemmas are used
in Propositions \ref{proposition1} and \ref{proposition2}.

\begin{lemma}\label{lemma3}
Under the same assumptions as in Lemma \ref{lemma1},
$\frac{1}{n-1}\sum_{t=1}^{n-1}\left(q\left(X_t;\widehat{\mathbf{\eta}}\right)-\sigma^2(X_t)\right)^2$
is consistent in the sense that:
\begin{itemize}
\item if $Q_1(n)\rightarrow 0$ as $n\rightarrow \infty$, then
\[ \frac{1}{n-1}\sum_{t=1}^{n-1}\left(q\left(X_t;\widehat{\mathbf{\eta}}\right)-\sigma^2(X_t)\right)^2\stackrel{p}{\longrightarrow} 0 \quad n\rightarrow
\infty;
\]
\item if, additionally, $Q_2(n)\rightarrow 0$ for some $\lambda>0$, then
\[
\frac{1}{n-1}\sum_{t=1}^{n-1}\left(q\left(X_t;\widehat{\mathbf{\eta}}\right)-\sigma^2(X_t)\right)^2\stackrel{a.s.}{\longrightarrow}
0 \quad n\rightarrow \infty.
\]
\end{itemize}
\end{lemma}
\noindent\textit{\textbf{Proof:}} By Theorem (3.2) in \cite{FraDia06}, which uses
the same line of the proof as in Theorem (16.1) in
\cite{Gyorfy02}, we have that \begin{displaymath}
W_1:=\sup_{f\in\mathcal{F}_n}\left|\frac{1}{n-1}\sum_{k=1}^{n-1}\left|f(X_k)-X_{k+1}^2\right|^2-E\left\{\left|f(X_t)-X_{t+1}^2\right|^2\right\}\right|\stackrel{p
(a.s.)}{\longrightarrow}0
\end{displaymath}
for $n\rightarrow \infty$. The above convergence is in probability or almost sure according
to the two conditions, $Q_1(n)\rightarrow 0$ and
$Q_2(n)\rightarrow 0$, respectively. Below, we use only the
convergence in probability because the almost sure convergence
follows exactly the same technique.

The neural network estimator
$q\left(X_t;\widehat{\mathbf{\eta}}\right)\in\mathcal{F}_n$ for some
$n>n_0$. Using model (\ref{eqn01}) we can write
\begin{eqnarray}\label{eqn01app} \nonumber
&&W_2:=\frac{1}{n-1}\sum_{k=1}^{n-1}\left(q\left(X_k;\widehat{\mathbf{\eta}}\right)-\sigma^2(X_k)\right)^2-E\left\{\left(q\left(X_t;\widehat{\mathbf{\eta}}\right)-\sigma^2(X_t)\right)^2\right\}+\\
\nonumber
&&+\frac{1}{n-1}\sum_{k=1}^{n-1}\left(\sigma^2(X_k)\varepsilon_{k+1}^2-\sigma^2(X_k)\right)^2-E\left\{\left(\sigma^2(X_t)\varepsilon_{t+1}^2-\sigma^2(X_t)\right)^2\right\}+\\
\nonumber
&&-\frac{2}{n-1}\sum_{k=1}^{n-1}\left|q\left(X_k;\widehat{\mathbf{\eta}}\right)-\sigma^2(X_k)\right|\left|\sigma^2(X_k)\varepsilon_{k+1}^2-\sigma^2(X_k)\right|+\\
&&+2E\left\{\left|q\left(X_k;\widehat{\mathbf{\eta}}\right)-\sigma^2(X_k)\right|\left|\sigma^2(X_k)\varepsilon_{k+1}^2-\sigma^2(X_k)\right|\right\}
\end{eqnarray}
Therefore $W_2\stackrel{p}{\longrightarrow}0$.
Consider the terms in the second line of (\ref{eqn01app}). By
assumptions (a) it follows that
$E\left\{\left(\sigma^2(X_t)\varepsilon_{t+1}^2-\sigma^2(X_t)\right)^2\right\}=E\left[\sigma^4(X_t)\right]E\left\{\left(\varepsilon^2_{t}-1\right)^2\right\}:=c_1$
with $0<c<\infty$.
By ergodicity of the process $\{X_t\}$ and using the assumptions
(a) we have that \begin{displaymath}
\frac{1}{n-1}\sum_{k=1}^{n-1}\left(\sigma^2(X_k)\varepsilon_{k+1}^2-\sigma^2(X_k)\right)^2\stackrel{a.s.}{\longrightarrow}c_1
\qquad n\rightarrow\infty.
\end{displaymath}
Since $E\left[\sigma^4(X_t)\right]<\infty$ then
$E\left\{\left(q\left(X_t;\widehat{\mathbf{\eta}}\right)-\sigma^2(X_t)\right)^2\right\}<\infty$
and by Lemma \ref{lemma1}
\[
E\left\{\left(q\left(X_t;\widehat{\mathbf{\eta}}\right)-\sigma^2(X_t)\right)^2\right\}\stackrel{p}{\longrightarrow}0
\]
when $n\rightarrow\infty$. Using Schwartz's inequality we have
that
\[E\left\{\left|q\left(X_k;\widehat{\mathbf{\eta}}\right)-\sigma^2(X_k)\right|\left|\sigma^2(X_k)\varepsilon_{k+1}^2-\sigma^2(X_k)\right|\right\}\stackrel{p}{\longrightarrow}0.
\]
So we have that
\begin{eqnarray*}
&&W_3:=\frac{1}{n-1}\sum_{k=1}^{n-1}\left(q\left(X_k;\widehat{\mathbf{\eta}}\right)-\sigma^2(X_k)\right)^2-\frac{2}{n-1}\sum_{k=1}^{n-1}\left|q\left(X_k;\widehat{\mathbf{\eta}}\right)-\sigma^2(X_k)\right|\left|\sigma^2(X_k)\varepsilon_{k+1}^2-\sigma^2(X_k)\right|
\end{eqnarray*}
and $W_3\stackrel{p}{\longrightarrow}0$.
Since
$E\left\{\left(q\left(X_t;\widehat{\mathbf{\eta}}\right)-\sigma^2(X_t)\right)^2\right\}<\infty$
then it follows that \begin{displaymath}
\frac{1}{n-1}\sum_{k=1}^{n-1}\left(q\left(X_k;\widehat{\mathbf{\eta}}\right)-\sigma^2(X_k)\right)^2\stackrel{p}{\longrightarrow}c_2
\end{displaymath}
when $n\rightarrow\infty$, with $0\le c_2<\infty$. So, it implies
that \begin{displaymath}
\frac{1}{n-1}\sum_{k=1}^{n-1}\left|q\left(X_k;\widehat{\mathbf{\eta}}\right)-\sigma^2(X_k)\right|\left|\sigma^2(X_k)\varepsilon_{k+1}^2-\sigma^2(X_k)\right|\stackrel{p}{\longrightarrow}c_2/2
\end{displaymath}
when $n\rightarrow\infty$. But, by Schwartz's inequality we can
write \begin{eqnarray*}
&&\frac{1}{n-1}\sum_{k=1}^{n-1}\left|q\left(X_k;\widehat{\mathbf{\eta}}\right)-\sigma^2(X_k)\right|\left|\sigma^2(X_k)\varepsilon_{k+1}^2-\sigma^2(X_k)\right|\le\\
&&\le\left[\frac{1}{n-1}\sum_{k=1}^{n-1}\left(q\left(X_k;\widehat{\mathbf{\eta}}\right)-\sigma^2(X_k)\right)^2\right]^{1/2}\left[\frac{1}{n-1}\sum_{k=1}^{n-1}\left(\sigma^2(X_k)\varepsilon_{k+1}^2-\sigma^2(X_k)\right)^2\right]^{1/2}.
\end{eqnarray*}
If we apply convergence, we have $c_2/2\le \sqrt{c_1c_2}$. Since $c_1$
can be considered an arbitrary constant because it depends on the
fourth moment of $\varepsilon_t$, while $c_2$ does not, the inequality is true if and only if $c_2=0$. This
completes the proof.\hfill $\Box$

\begin{lemma}\label{lemma4}
Under the same assumptions as in Lemma \ref{lemma1},
$\frac{1}{n-1}\sum_{t=1}^{n-1}\left(q^{(2)}\left(X_t;\widehat{\mathbf{\eta}}\right)-\sigma^2_{(2)}(X_t)\right)^2$
is consistent in the sense that:
\begin{itemize}
\item if $Q_1(n)\rightarrow 0$ as $n\rightarrow \infty$, then
\[ \frac{1}{n-1}\sum_{t=1}^{n-1}\left(q^{(2)}\left(X_t;\widehat{\mathbf{\eta}}\right)-\sigma^2_{(2)}(X_t)\right)^2\stackrel{p}{\longrightarrow} 0 \quad n\rightarrow
\infty;
\]
\item if, in addition, $Q_2(n)\rightarrow 0$ for some $\lambda>0$, then
\[
\frac{1}{n-1}\sum_{t=1}^{n-1}\left(q^{(2)}\left(X_t;\widehat{\mathbf{\eta}}\right)-\sigma^2_{(2)}(X_t)\right)^2\stackrel{a.s.}{\longrightarrow}
0 \quad n\rightarrow \infty.
\]
\end{itemize}
\end{lemma}
\noindent\textit{\textbf{Proof:}}
As in the proof of Lemma (2), let $\mathcal{G}$ be the class of
all functions $\sigma^2(x)$ which satisfy the assumptions (a2) and
(a3). Now, we have that
\begin{displaymath}
\frac{1}{n-1}\sum_{k=1}^{n-1}\left(q^{(2)}\left(X_t;\widehat{\mathbf{\eta}}\right)-\sigma^2_{(2)}(X_t)\right)^2\le\left\|\widehat{d}^2_n\right\|^2\frac{1}{n-1}\sum_{k=1}^{n-1}\left(q\left(X_t;\widehat{\mathbf{\eta}}\right)-[\sigma^2(X_t)]\right)^2
\end{displaymath}
where
$\left\|d^2_n\right\|^2=\sup_{f\in\mathcal{G}}\frac{1}{n-1}\sum_{k=1}^{n-1}\left[f''(x_k)\right]^2$
and
$\left\|\widehat{d}^2_n\right\|^2=\sup_{f\in\mathcal{G}}\frac{1}{n-1}\sum_{k=1}^{n-1}\left[f''(X_k)\right]^2$,
with the stochastic process $\{X_t\}$ defined in (\ref{eqn01}) and
$\left\|\cdot\right\|$ the norm of $L_2$ space with respect to the
empirical measure.
Based on assumption (a3), every function in $\mathcal{G}$ has a bounded
second derivative and so \begin{displaymath}
\lim_{n\rightarrow\infty}\frac{1}{n-1}\sum_{k=1}^{n-1}\left[f''(x_k)\right]^2<\infty
\end{displaymath}
for every sequence $\{x_k\}\in\mathbb{R}$, $k=1,2,\ldots$.

Based on assumptions (a) and ergodicity of the stochastic process
$\{X_t\}$ we have that
$\left\|\widehat{d}^2_n\right\|^2\stackrel{a.s.}{\longrightarrow}c<\infty$.
Finally, using Lemma (3) it follows that \begin{displaymath}
\left\|\widehat{d}^2_n\right\|^2\frac{1}{n-1}\sum_{k=1}^{n-1}\left(q\left(X_t;\widehat{\mathbf{\eta}}\right)-\sigma^2(X_t)\right)^2\stackrel{p(a.s.)}{\longrightarrow}0
\qquad n\rightarrow\infty.
\end{displaymath}
The above convergence is in probability if $Q_1(n)\rightarrow 0$,
when $n\rightarrow\infty$. If, in addition, $Q_2(n)\rightarrow 0$,
when $n\rightarrow\infty$, then there will be almost sure convergence.
This completes the proof.\hfill $\Box$\\

\subsection{Consistency for the Functional of the bias}
 
Let $I_x=[x-a/2,x+a/2]$, with $a>0$ for all $x\in\mathbb{R}$.
According to assumption (a5) it follows that $\mu_X(I_x)>0$.
Moreover, the number of observed values in $I_x$ from
(\ref{eqn01}) tends to infinity when $n\rightarrow\infty$ with
probability 1.

Using model (\ref{eqn01}),
we can write the functional of the bias, $\mathbb{B}_{\omega_{I_x}}$, as
\begin{equation}\label{fun1app}
\mathbb{B}_{\omega_{I_x}}=
C_1^2\int_{I_x}\left(\sigma^2_{(2)}(x)\right)^2f_X(x)d\omega_{I_x}.
\end{equation}
Similarly, we can write its estimator as
$\widehat{\mathbb{B}}_{\omega_{I_x}}$, that is
\begin{equation}\label{Rhatapp}
\widehat{\mathbb{B}}_{\omega_{I_x}}=\frac{C_1^2\sum_{t=1}^{n-1}\left[q^{(2)}\left(X_t;\widehat{\mathbf{\eta}}\right)\right]^2\mathbb{I}(X_t\in
I_x)}{\sum_{t=1}^{n-1}\mathbb{I}(X_t\in I_x)}
\end{equation}
as reported in  section 1.2 of this Supplement.
\begin{proposition}\label{proposition1}
Under the same assumptions as in Lemma \ref{lemma1}, $\widehat{\mathbb{B}}_{\omega_{I_x}}$, defined in
(\ref{Rhatapp}), is consistent in the
sense that:
\begin{itemize}
\item If $Q_1(n)\rightarrow 0$ as $n\rightarrow \infty$, then
\[ \widehat{\mathbb{B}}_{\omega_{I_x}}\stackrel{p}{\longrightarrow} \mathbb{B}_{\omega_{I_x}} \quad n\rightarrow
\infty
\]
\item if, additionally, $Q_2(n)\rightarrow 0$ for some $\lambda>0$, then
\[
\widehat{\mathbb{B}}_{\omega_{I_x}}\stackrel{a.s.}{\longrightarrow}
\mathbb{B}_{\omega_{I_x}} \quad n\rightarrow \infty
\]
\end{itemize}
where $\mathbb{B}_{\omega_{I_x}}$ is defined in (\ref{fun1app}).
\end{proposition}

\vspace{20pt}\noindent\textit{\textbf{Proof:}} For the sake of simplicity we consider only the convergence in probability. The almost sure convergence follows the same technique.
The estimator in (\ref{Rhatapp}) can be written as
\[
\widehat{\mathbb{B}}_{\omega_{I_x}}=\frac{C_1^2\frac{1}{n-1}\sum_{t=1}^{n-1}\left[q^{(2)}\left(X_t;\widehat{\mathbf{\eta}}\right)\right]^2\mathbb{I}(X_t\in
I_x)}{\frac{1}{n-1}\sum_{t=1}^{n-1}\mathbb{I}(X_t\in I_x)}.
\]
The quantity $C_1^2$ is known. By ergodicity of the stochastic
process $\{X_t\}$ it follows that
\begin{displaymath}
\frac{1}{n-1}\sum_{t=1}^{n-1}\mathbb{I}(X_t\in
I_x)\stackrel{a.s.}{\longrightarrow}\mu_X(I_x).
\end{displaymath}
Using assumptions (a) and again ergodicity of the stochastic
process $\{X_t\}$ we have that
\begin{displaymath}
\frac{1}{n-1}\sum_{t=1}^{n-1}\left(\sigma^2_{(2)}(X_t)\right)^2\mathbb{I}(X_t\in
I_x)\stackrel{a.s.}{\longrightarrow}\int_{I_x}\left(\sigma^2_{(2)}(x)\right)^2f_X(x)dx.
\end{displaymath}
By Lemma \ref{lemma4}
$\frac{1}{n-1}\sum_{t=1}^{n-1}\left[q^{(2)}\left(X_t;\widehat{\mathbf{\eta}}\right)\right]^2$
and
$\frac{1}{n-1}\sum_{t=1}^{n-1}\left(\sigma^2_{(2)}(X_t)\right)^2$
converge in probability to the same limit. But the result is the
same if we consider
\[\frac{1}{n-1}\sum_{t=1}^{n-1}\left[q^{(2)}\left(X_t;\widehat{\mathbf{\eta}}\right)\right]^2\mathbb{I}(X_t\in
I_x).\] Therefore, it follows that
\begin{displaymath}
\frac{1}{n-1}\sum_{t=1}^{n-1}\left[q^{(2)}\left(X_t;\widehat{\mathbf{\eta}}\right)\right]^2\mathbb{I}(X_t\in
I_x)\stackrel{p}{\longrightarrow}\int_{I_x}\left(\sigma^2_{(2)}(x)\right)^2f_X(x)dx
\end{displaymath}
Since $d\omega_{I_x}g=dx/\mu_X(I_x)$ and $\mu_X(I_x)>0$ we have that
$\widehat{\mathbb{B}}_{\omega_{I_x}}\stackrel{p}{\longrightarrow}\mathbb{B}_{\omega_{I_x}}$.
The proof is complete.\hfill $\Box$\\

Let $m_{4\varepsilon}=E(\varepsilon_t^4)$ and $\widehat{m}_{4\varepsilon}=\frac{\sum_{t=2}^{n} X_{t-1}^4}{\sum_{t=2}^{n} [q(X_{t-1}, \B{\widehat{ \eta}})]^2}$.
\begin{corollary}\label{corollary2}
Using the same conditions as in Proposition \ref{proposition1},
the estimator $\widehat{m}_{4\varepsilon}$, is consistent in the sense that:
\begin{itemize}
\item if $Q_1(n)\rightarrow 0$ as $n\rightarrow \infty$, then
$\widehat{m}_{4\varepsilon}\stackrel{p}{\longrightarrow} m_{4\varepsilon} \quad n\rightarrow
\infty$
\item if, in addition, $Q_2(n)\rightarrow 0$ for some
$\lambda>0$, then
$\widehat{m}_{4\varepsilon}\stackrel{a.s.}{\longrightarrow}
m_{4\varepsilon} \quad n\rightarrow \infty$.
\end{itemize}
\end{corollary}
\noindent\textit{\textbf{Proof:}}
As in the previous proofs, we analyze the convergence in
probability since the almost sure convergence is straightforward.
The estimator
$\widehat{m}_{4\varepsilon}$ can be written as
\begin{displaymath}
\widehat{ m}_{4\varepsilon} = \frac{\frac{1}{n-1}\sum_{t=1}^{n-1}
X_t^4}{\frac{1}{n-1}\sum_{t=1}^{n-1}
\left[q\left(X_t;\widehat{\mathbf{\eta}}\right)\right]^2}.
\end{displaymath}
Based on assumptions (a) and ergodicity of the stochastic process
$\{X_t\}$, it follows that
\begin{displaymath}
\frac{1}{n-1}\sum_{t=1}^{n-1}X_t^4\stackrel{a.s.}{\longrightarrow}E\left[\sigma^4(X_t)\right]m_{4\varepsilon}\qquad
n\rightarrow\infty
\end{displaymath}
and
\begin{displaymath}
\frac{1}{n-1}\sum_{t=1}^{n-1}\sigma^4(X_t)\stackrel{a.s.}{\longrightarrow}E\left[\sigma^4(X_t)\right]\qquad
n\rightarrow\infty
\end{displaymath}
since $E\left[\sigma^4(X_t)\right]<\infty$ and
$m_{4\varepsilon}<\infty$. Using Lemma (3), it implies that
$\frac{1}{n-1}\sum_{t=1}^{n-1}\sigma^4(X_t)$ and
$\frac{1}{n-1}\sum_{t=1}^{n-1}
\left[q\left(X_t;\widehat{\mathbf{\eta}}\right)\right]^2$ have the
same limit in probability. Therefore,
\[\frac{1}{n-1}\sum_{t=1}^{n-1}
\left[q\left(X_t;\widehat{\mathbf{\eta}}\right)\right]^2\stackrel{p}{\longrightarrow}E\left[\sigma^4(X_t)\right],
\]
when $n\rightarrow\infty$. We can conclude that
$\widehat{m}_{4\varepsilon}\stackrel{p}{\longrightarrow}m_{4\varepsilon}$.
The proof is complete.\hfill $\Box$\\

\subsection{Consistency for the functional of variance}

Using model (\ref{eqn01}), we can write the functional of variance, $\mathbb{V}_{\omega_{I_x}}$, as
\begin{equation}\label{fun1app1}
\mathbb{V}_{\omega_{I_x}}=
C_2\int_{I_x}\left[\sigma^4(u)\right]d\omega_{I_x}(u)
\left(m_{4\varepsilon}-1\right).
\end{equation}
Similarly, we can write its estimator as
\begin{equation}\label{Rhatapp1}
\widehat{\mathbb{V}}_{\omega_{I_x}}=\frac{C_2\sum_{i=1}^{n^*}\left[q\left(z_i;\widehat{\mathbf{\eta}}\right)\right]^2/n^*}{\sum_{t=1}^{n}\mathbb{I}(X_t\in
I_x)/n} \left(\widehat{m}_{4\varepsilon}-1\right).
\end{equation}
as reported in section 1.2 of this Supplement.
The points $\{z_1, z_2, \ldots, z_{n^*}\}$ are uniformly spaced
values from the interval $I_x$, with $n^*=O(n)$.

\begin{proposition}\label{proposition2}
Using the same conditions
as in Proposition \ref{proposition1}, then $\widehat{\mathbb{V}}_{\omega_{I_x}}$, defined
in (\ref{Rhatapp1}), with $I_x\subset\mathbb{R}$ and $n^*=O(n)$,
is consistent in the sense that:
\begin{itemize}
\item If $Q_1(n)\rightarrow 0$ as $n\rightarrow \infty$, then
\[
\widehat{\mathbb{V}}_{\omega_{I_x}}\stackrel{p}{\longrightarrow}
\mathbb{V}_{\omega_{I_x}} \quad n\rightarrow \infty
\]
\item if, in addition, $Q_2(n)\rightarrow 0$ for some
$\lambda>0$, then
\[
\widehat{\mathbb{V}}_{\omega_{I_x}}\stackrel{a.s.}{\longrightarrow}
\mathbb{V}_{\omega_{I_x}} \quad n\rightarrow \infty.
\]
\end{itemize}
\end{proposition}
\noindent\textit{\textbf{Proof:}}
As in the previous proofs, we analyze the convergence in
probability since the almost sure convergence is straightforward.\\ By
Corollary \ref{corollary2}, it follows that
$\widehat{m}_{4\varepsilon}\stackrel{p}{\longrightarrow}m_{4\varepsilon}$,
 when $n\rightarrow\infty$. Using the ergodicity of the stochastic
process $\{X_t\}$, we have that $\sum_{t=1}^{n}\mathbb{I}(X_t\in
I_x)/n\stackrel{a.s.}{\longrightarrow}\mu_X(I_x)$, when
$n\rightarrow\infty$.\\ Since $C_2$ is a known quantity, we need
only to show that \begin{displaymath}
\sum_{i=1}^{n^*}\left[q\left(z_i;\widehat{\mathbf{\eta}}\right)\right]^2/n^*\stackrel{p}{\longrightarrow}\int_{I_x}\left[\sigma^4(u)\right]du
\end{displaymath}
where the points $\{z_1, z_2, \ldots, z_{n^*}\}$ are deterministic
and  uniformly spaced values from the interval $I_x$.

By Lemma \ref{lemma1} and Lemma \ref{lemma3}, we know that
\begin{equation}\label{distapp}
\inf_{s\in\mathcal{F}_n}\int_{\mathbb{R}}\left(s(x)-\sigma^2(x)\right)^2d\mu_X(x)\rightarrow
0\qquad n\rightarrow\infty
\end{equation}
\begin{equation}\label{varapp}
\sup_{s\in\mathcal{F}_n}\left|\sum_{k=1}^{n-1}\left|s(X_k)-X^2_{k+1}\right|^2-E\left\{\left|s(X_t)-X_{t+1}^2\right|^2\right\}\right|\stackrel{p}{\longrightarrow}0\qquad
n\rightarrow\infty.
\end{equation}
Both (\ref{distapp}) and (\ref{varapp}) refer to the Neural
Network estimator with respect to the stochastic process
$\{X_t\}$. Instead, we consider some points which are not drawn by
the process. So, in this case, we have to show that
(\ref{distapp}) and (\ref{varapp}) hold.

By assumption (a5), we have that
\begin{displaymath}
\inf_{s\in\mathcal{F}_n}\int_{\mathbb{R}}\left(s(x)-\sigma^2(x)\right)^2d\mu_X(x)\ge
\inf_{s\in\mathcal{F}_n}\int_{I_x}\left(s(x)-\sigma^2(x)\right)^2f_X(x)dx\ge
\end{displaymath}
\begin{displaymath}
\ge
C_f\inf_{s\in\mathcal{F}_n}\int_{I_x}\left(s(x)-\sigma^2(x)\right)^2dx
\end{displaymath}
where $C_f:=\min_{x\in I_x}\{f_X(x)\}$, with $0<C_f<\infty$. By
(\ref{distapp}), it follows that
\begin{displaymath}
\inf_{s\in\mathcal{F}_n}\int_{I_x}\left(s(x)-\sigma^2(x)\right)^2dx\rightarrow
0\qquad n\rightarrow\infty.
\end{displaymath}
Thus, we have proved that (\ref{distapp}) is also true with
respect to the points $\{z_1, z_2, \ldots, z_{n^*}\}$ uniformly
spaced from the interval $I_x$. Since $n^*=O(n)$, we can consider
asymptotically $n$ instead of $n^*$.

Put $z_i=\widetilde{X}_i+(z_i-\widetilde{X}_i)$, where
$\widetilde{X}_i\in\{X_1,X_2,\ldots,X_n\}$, $i=1,2,\ldots,n^*$. For
every $\epsilon>0$ and $z_i$, we choose a $\widetilde{X}_i$ such that
$\left|z_i-\widetilde{X}_i\right|<\epsilon$. Now, we have to show
that such a $\widetilde{X}_i$ exists with probability 1.

By assumption (a5) and based on Proposition (A1.7) in
\cite{Ton90}, every non null compact set is a ``small set" with
respect to the Lebesgue measure for the Markov process in
(\ref{eqn01}). But every set of radius $\epsilon$, which contains
$z_i$ is non null compact set using the Lebesgue measure.
Therefore, there exists a $n_0$ such that for each $n>n_0$ we can
find at least a $\widetilde{X}_i\in\{X_1,X_2,\ldots,X_n\}$ with
probability 1.\\ Define $d_i:=(z_i-\widetilde{X}_i)$. Then
$|d_i|<\infty$ with probability 1, when $n\rightarrow\infty$,
$\forall i$.\\ Define $Z_i:=(\widetilde{X}_i,d_i)$. The bi-dimensional
random variables $Z_i$ retain the property of exponentially
$\alpha$-mixing because we have only deterministic variables $z_i$
and random variables $X_i$ which are exponentially
$\alpha$-mixing. Since $n^*=O(n)$, we can write, asymptotically,
\begin{eqnarray*}
&& \sup_{s\in\mathcal{F}_n}\left|\sum_{k=1}^{n-1}\left|s(X_k)-X^2_{k+1}\right|^2-E\left\{\left|s(X_t)-X_{t+1}^2\right|^2\right\}\right| \\
&\le&
\sup_{s\in\mathcal{F}_n}\left|\sum_{k=1}^{n-1}\left|s(Z_k)-\widetilde{X}^2_{k+1}\right|^2-E\left\{\left|s(Z_t)-\widetilde{X}_{t+1}^2\right|^2\right\}\right|
\end{eqnarray*}
because, using the proof of Theorem (3.2) from \cite{FraDia06}, the
upper bounds for the \textit{sup} depend on $d_n$, $\Delta_n$ and
the dimension of the input variables. But this dimension is 1 in
(\ref{varapp}) and 2 if we use $Z_i$ as input variables, that is
the uniformly spaced values in $I_x$.\\ Therefore, these upper
bounds are the same when $n\rightarrow\infty$. So, it follows that
\begin{displaymath}
\sup_{s\in\mathcal{F}_n}\left|\sum_{k=1}^{n-1}\left|s(Z_k)-\widetilde{X}^2_{k+1}\right|^2-E\left\{\left|s(Z_t)-\widetilde{X}_{t+1}^2\right|^2\right\}\right|\stackrel{p}{\longrightarrow}0\qquad
n\rightarrow\infty
\end{displaymath}
In this way, we can apply Lemma (3) in the case of the uniformly
spaced values in $I_x$. Then we have that
$\sum_{i=1}^{n^*}\left[q\left(z_i;\widehat{\mathbf{\eta}}\right)\right]^2/n^*$
and $\sum_{i=1}^{n-1}\left(\sigma^4(z_i)\right)/n$ have the same
limit in probability, when $n\rightarrow\infty$.\\ Finally, the
result follows.\hfill $\Box$\\

%%%%%%%%%%%%%%%%%%%%%%%%%%%%%%%%%%%%%%%%%%%%%%%%%%%%%%%%%%%%%%%%%%%%%%%%%%%%%%%%%%%%%%%%%%%%%%%%%%%%%%%%%%%%%%%%%%%%%%%%%%%%
\vskip 14pt
\noindent {\textbf{Acknowledgements}}
Financial support by the Italian Ministry of Education, University and Research (MIUR), PRIN Research Project 2010--2011 -- prot.
2010J3LZEN, ``Forecasting economic and financial time series:
understanding the complexity and modelling structural change'', is
gratefully acknowledge.

\end{document}